\documentclass[manuscript]{aastex}
\usepackage{graphicx}
\usepackage{epstopdf}
\usepackage{xfrac}
\usepackage{subfigure}
\usepackage{amsmath}

\shorttitle{Charge Exchange Induced X-ray Emission of Fe~XXV and Fe~XXVI via a Streamlined Model}
\shortauthors{Mullen et al.}

\begin{document}

\title{Charge Exchange Induced X-ray Emission of Fe~XXV and Fe~XXVI via a Streamlined Model}
\author{P. D. Mullen, R. S. Cumbee, D. Lyons, and P. C. Stancil}
\affil{Department of Physics and Astronomy and the Center for Simulational Physics, University of Georgia, Athens, GA 30602}

\begin{abstract}
Charge exchange is an important process for the modeling of X-ray spectra obtained by the \textit{Chandra}, \textit{XMM-Newton}, and \textit{Suzaku} X-ray observatories, as well as the anticipated \textit{Astro-H} mission. The understanding of the observed X-ray spectra produced by many astrophysical environments is hindered by the current incompleteness of available atomic and molecular data -- especially for charge exchange.  Here, we implement a streamlined program set that applies quantum defect methods and the Landau-Zener theory to generate total, $n$-resolved, and $n\ell S$-resolved cross sections for any given projectile ion/ target charge exchange collision.  Using this data in a cascade model for X-ray emission, theoretical spectra for such systems can be predicted.  With these techniques, Fe$^{25+}$ and Fe$^{26+}$ charge exchange collisions with H, He, H$_{2}$, N$_{2}$, H$_2$O, and CO are studied for single electron capture.  These systems have been selected as they illustrate computational difficulties for high projectile charges.  Further, Fe XXV and Fe XXVI emission lines have been detected in the Galactic center and Galactic ridge.   Theoretical X-ray spectra for these collision systems are compared to experimental data generated by an electron beam ion trap study.  Several $\ell$-distribution models have been tested for Fe$^{25+}$ and Fe$^{26+}$ single electron capture.  Such analysis suggests that commonly used $\ell$-distribution models struggle to accurately reflect the true distribution of electron capture as understood by more advanced theoretical methods.  
\end{abstract}

\section{Introduction}
 The charge exchange (CX) process may be a large contributor to the X-ray emission of many astrophysical environments such as comets, supernova remnants, the heliosphere, astrospheres of stars, and generally, highly ionized regions of the interstellar medium \citep[e.g.,][]{lis96,cra00,6,war2,bha07,10, suda}.  In modeling CX emission, the availability of atomic data for charge exchange is often insufficient thus hindering the completeness and validity of present models.  Therefore, working at the interface of atomic and molecular physics with astrophysics, we have applied quantum defect methods, the Landau-Zener theory of charge exchange, and a cascade model for X-ray emission to show that not only can charge exchange calculations be performed, but that they must be considered in the X-ray spectroscopy of many astrophysical environments.  \\
\indent Charge exchange collisions occur when a projectile ion captures an electron from a target neutral species.   In this work, we consider only single electron capture (SEC). This process is given generally by Equation (1) where $X$ denotes the projectile ion with charge $q$ and $Y$ denotes the target species, 
\begin{equation}
X^{q+} + Y \longrightarrow X^{(q-1)+} \: (n\ell \; ^{2S+1}L) \: + Y^{+} + \Delta E, 
\end{equation}
while $n$ and $\ell$ are the principal and orbital angular momentum of the captured electron, and $S$ and $L$ are the total spin and total orbital electron angular momenta of the product ion. 
\linebreak
\indent The multi-channel Landau-Zener (MCLZ) approach is applied to generate $n$-resolved, $n\ell S$-resolved, and total cross sections for charge exchange.  The MCLZ level of theory is not the most advanced framework as compared to other methods such as quantum-mechanical molecular-orbital close-coupling (QMOCC)
\citep[e.g.,][]{6}.  However, the power of MCLZ calculations can be seen in its quick computation time --relative to other techniques-- and the implications and quick resolution that such a tool could bring to the current lack of atomic and molecular data necessary for the astrophysical modeling for many systems.  Here, we also validate our theoretical spectra for Fe XXV and Fe XXVI with molecular nitrogen charge exchange collisions by comparison to experimental spectra given in \cite{1, war2}. Atomic units are used throughout, unless indicated otherwise. 
\section{Theory}
\subsection{Landau-Zener Approximations}
The foundation for the multi-channel Landau-Zener theory begins by examining the initial and final channels of the collision system.  The initial channel is characterized by the interaction between the projectile ion, with charge $q_1$, and target species prior to collision.  This interaction is modeled by Equation (2) where the potential energy of the channel is given as a function of internuclear distance $R$, 
\begin{equation}
V_i = A\exp(-BR) - \frac{\alpha q_1^2}{2R^4},
\end{equation}
where coefficients $A$ and $B$ are estimated in \cite{2}.  The second term represents a polarization interaction between the projectile and target, with $\alpha$ being the dipole polarizability of the neutral target, whereas the first term accounts for the repulsive nature of the system at small internuclear distances.  
\linebreak
\indent The final channel $f$ of the collision system is a Coulombic repulsion interaction because the two product species are of positive charge.  Therefore, a simple Coulomb repulsion potential is used, 
\begin{equation}
V_f = \frac{(q_1-1)(q_2)}{R} -\Delta E_\infty, 
\end{equation}
where, $(q_1-1)$ is the charge of the projectile after the collision and $q_2$ is the final charge of the target.  For SEC between an ion and neutral, $q_2 = +1$.   The $\Delta E_\infty$ term is the energy released as a result of the charge exchange process and is precisely the sum of 1) the difference in ionization potentials of the product ion and the target species, and 2) the excitation energy of the product ion.
\linebreak
\indent
The charge exchange probability is defined in terms of the transition from the initial channel to the final channel.  The internuclear distance at which this transition can occur is referred to as the avoided crossing, $R_c$.  The avoided crossing distances are estimated by setting the initial and final potential curves equal to one another.  These avoided crossings are used as input parameters in calculating the energy difference, $\Delta V$, of the adiabatic potential at $R_c$.  We estimate $\Delta V$ using the Olson-Salop-Tauljberg adaptation \citep{other, 3, 4} displayed in Equation (4):
\begin{equation}
\Delta V = (9.13 f_{n \ell} / \sqrt{q_1}) \exp (-1.324 R_c \beta  \sqrt{q_1}), 
\end{equation}
where $\beta$, $q_1$, and $f_{n \ell}$ are detailed in \cite{4}.
\linebreak
\indent The charge exchange probability is thus dependent on three parameters: $R_c$, $\Delta V$, and the absolute value of the difference in slopes between the initial and final potential curves, $\mid V'_f - V'_i \mid$.  For a simple, 2-channel system, the charge exchange probability is given as,  
\begin{equation}
\varphi_{if} = 2p_{if}(1-p_{if}), 
\end{equation}
where $p_{if}$ is given in \citet{2} as
\begin{equation}
p_{if} = \exp(-\omega_{if}), 
\end{equation}
\begin{equation}
\omega_{if} = \frac{4 \pi^2 \Delta V^2}{hv_{R_c} \mid V'_f - V'_i \mid_{R_c}}, 
\end{equation}
\linebreak
and $v_{R_c}$ is the radial component of the relative velocity of the collision \citep{2}. \\
\indent  In this work, we compute multi-channel collision probabilities using the \cite{5} relation for all allowed product ion states. The possible product states are determined following the Wigner-Witmer rules \citep{herz,wigner} for the angular momenta of the molecular channels and for channels connected to the initial channel via nonadiabatic radial coupling.  Therefore, this probability computation becomes complicated with increasing number of product ion states, which increases with $q_1$, and may result in long computation times. 
\linebreak
Finally, with these probabilities, we compute cross sections for charge exchange by considering the probability, collision energy, and all partial waves, $J$, of the collision system, 
\begin{equation}
\sigma_{if} = \frac{\pi}{k_i^2} \sum\limits_{J=0}^{J_{\rm max}} (2J + 1)\varphi^J_{if} ,
\end{equation}
where $k_i$ is the wave vector for relative motion in the initial channel $i$.
\subsection{$\ell$-Distribution Models}
For bare ion collision systems, the product ion with a newly captured electron has $\ell$-states that are degenerate within a given $n$-state.  The Landau-Zener theory requires that avoided crossing distances are well separated.  Therefore, we must apply $\ell$-distribution models to MCLZ $n$-resolved cross sections for bare ion collision systems.  
\linebreak
\indent The \textit{statistical} $\ell$-distribution discussed in \cite{6} is given by, 
\begin{equation}
W^{\rm{Stat}}_{n\ell} = (2 \ell + 1)/ n^{2},
\end{equation}
and is recommended for collision energies $\gtrsim$ 10 keV/u.  
The \textit{low energy} $\ell$-distribution \citep[][and references therein]{6} is given by,
\begin{equation}
W^{\rm{Low}}_{n\ell} = (2 \ell + 1) \frac{[(n-1)!]^{2}}{(n+\ell)!(n-1-\ell)!},
\end{equation}
and is recommended for collision energies $\lesssim$ 10-100 eV/u.  Two other $\ell$-distribution models, obtained from and adopted in the AtomDB Charge eXchange \citep[ACX,][]{7} model, are tested.  We refer to them as the \textit{separable} distribution, as displayed in Equation (11), and the \textit{modified low energy} distribution given in Equation (12),
\begin{equation}
W_{n\ell}^{\rm{Sep}} = \bigg(\frac{2\ell +1}{q_1} \bigg) \exp \bigg[\frac{- \ell(\ell+1)}{q_1} \bigg]
\end{equation}
and 
\begin{equation}
W_{n\ell}^{\rm{LowMod}} = \ell(\ell +1)(2\ell+1)\frac{(n-1)!(n-2)!}{(n+\ell)!(n-\ell-1)!},
\end{equation}
respectively. It is interesting to note that
Equation (12) gives a zero population for $\ell = 0$ ($s$) states and that Equation (11) is the only distribution dependent on incident ion charge. None of the
distribution functions have an explicit collision energy dependence.

\subsection{Cascade-Model for X-Ray Emission}
 After charge exchange, the product ion is unstable as the electron is captured into a highly excited state.  Therefore, the charged ion stabilizes by emitting a photon.   The photon energies associated with such cascading events range from the extreme UV to the X-ray portion of the electromagnetic spectrum since the largest radiative probabilities correspond to decay to the ground state.  Using $n\ell S$-resolved cross sections, the cascade model for X-ray emission assumes that atomic levels are populated based on their relative cross sections for electron capture into associated states. Using this data alongside transition probabilities, the cascade model tracks the evolution of the electron's cascade from its highly-energetic state to stability by considering every possible transition that obeys the dipole selection rules 
$
\Delta \ell = \pm 1
$
and
$
\Delta S = 0
$.  
Forbidden and intercombination transitions from the triplet $n=2$ levels of He-like ions are also included. 
By considering all possible routes for this cascade and the initial population of states, one can compute theoretical emission lines for charge exchange induced X-ray emission, assuming an optically thin plasma, following the approach of \citet{rig02}.  In this paper, we compare such theoretical spectra for Fe~XXV and Fe~XXVI, that utilize the MCLZ $n \ell S$-resolved cross sections, to the experimental spectra obtained with the electron beam ion trap (EBIT) experiment of \cite{1, war2}.  

\subsection{Quantum Defect Theory}
 It is essential to consider all possible final channels of a given collision system and to have detailed knowledge of their excitation energies in order to properly identify avoided crossing locations, and ultimately, the \cite{5} probability relations.  Therefore, excitation energies are gathered in order to determine the $\Delta E_\infty$ values pertinent in the avoided crossing estimates.  However, the main source for these excitation energies, the NIST Atomic Spectra Database (\citealt{9}), is often missing states -- especially high-lying Rydberg levels.  Therefore, we have employed quantum defect theory, which basically modifies the Rydberg formula through an introduced quantum defect, $\mu$.  In an effort to approximate excitation energies for all states that must be considered for charge exchange to create He-like Fe$^{24+}$, we have devised a strategy to establish trends in quantum defect parameters.  With these trends, quantum defect theory estimates energy levels via the modified Rydberg formula \citep[see for example,][]{8}, Equation (13), where the parameters of the calculation are the following: ionization potential  ($E_0$), $n$-level, $\ell$-level, and $\mu_{\ell}$.  Here,  $\mu_{\ell}$ also depends on the parameter $\delta$ resulting in the relations
\begin{equation}
E_{n\ell} = E_0\bigg(1-\frac{1}{n^{*2}}\bigg),
\end{equation}
where
\begin{equation}
n^{*} = n- \mu_{\ell}, 
\end{equation}
and
\begin{equation}
\mu_{\ell} = \ell + \frac{1}{2} - \biggl [ \biggl (\ell + \frac{1}{2} \biggr )^{2} + \delta \biggr ]^{\frac{1}{2}}.
\end{equation}
The strategy employed in estimating the unknown energy levels is by finding trends in the parameter $\delta$ from available data. We apply these trends to high-lying Rydberg-levels to estimate $\delta$ and ultimately, $E_{n \ell}$. 
\section{Results and Discussion} 
 Until this point, we have generalized the method to compute the charge exchange cross sections and resulting X-ray spectra for a given CX collision system.   Now, we will display the results in applying this methodology to the two exemplary cases of Fe$^{25+}$ and Fe$^{26+}$ CX collisions. These charge exchange calculations were chosen for three reasons.   First, the high charge of the Fe$^{25+}$ and Fe$^{26+}$ systems (i.e. $q=25$ and $q=26$, respectively) presents computational difficulties as more final channels (final product ion states) must be considered for the collision system.  The dominant $n$-capture channel is given approximately by $n_{\textrm{max}} \sim q^{0.75}$; within each $n$ level, the number of $\ell$ levels is equivalent to $n$.  Because MCLZ calculations typically include all product ion states from n=1 to $n_{\textrm{max}}$+3, the total number of channels roughly scales as $\sim n^2$ or increases with charge as $\sim q^{1.5}$. This results in the charge exchange probability expression becoming much more complicated, as seen in the \cite{5} probability relations, and ultimately stress tests the MCLZ code.  Such a calculation using a more advanced method (e.g., QMOCC) would require more than 100 molecular channels and 1000s of nonadiabatic coupling elements; such a calculation is not currently feasible.  Second, the presence of experimental X-ray emission data for these systems made these collisions an attractive option as it presents an opportunity to compare theoretical spectra to experiment.  Finally, the astrophysical relevance of Fe XXV and Fe XXVI emission in  the Galactic center and ridge \citep{tan99,mun04} makes these calculations useful for astrophysical modelers of these regions as charge exchange may be a dominant process. 
 \linebreak
 \indent Therefore, it is with this motivation that total, $n$-resolved, $\ell$-resolved, and $S$-resolved cross sections were calculated for Fe$^{25+}$ and Fe$^{26+}$ collisions with H, He, H$_2$, N$_2$, H$_2$O, and CO.
\subsection{Fe XXVI}
To begin studying Fe$^{26+}$ bare ion collisions, Wigner-Witmer rules are applied to determine possible product channels for the collision.  These rules, based on electronic angular momentum conservation principles, indicate that the initial channel for Fe$^{26+}$ colliding with H is a $^2\Sigma^+$, while for all other targets it is a $^1\Sigma^+$ state\footnote{For molecular targets we consider only linear geometries and do not resolve
the rotational or vibrational motion of the initial neutral or product molecular ion.}. All product Fe$^{25+}$ ions are doublets with a total electron spin $S=1/2$ (which we drop in the notation for Fe XXVI).  With this information we perform Landau-Zener approximations to obtain total, $n$-resolved, and $n\ell$-resolved cross sections.  First, we examine the total cross sections for the collision systems of Fe$^{26+}$ with various targets in Figure 1. One of these collision systems alone would take a significant amount of CPU months to calculate with more complex theories such as QMOCC -- and at only one collision energy.  Because of this effective Landau-Zener technique, total cross sections for Fe$^{26+}$ collisions with H, He, H$_2$, N$_2$, H$_2$O, and CO -- for collision energies from 0.001 eV/u to 100 keV/u -- are computed and quickly show an interesting trend that with increasing ionization potential of the target, we see decreasing total cross sections for collision energies $\gtrsim$ 10 eV/u. 
\linebreak
\indent
In addition to total cross sections, we can also investigate $n$-resolved cross sections.  As shown in Figure 2, MCLZ calculations show that the electron capture into $n$ = 12, 13, and 14 generate the largest cross sections for Fe$^{26+}$ collisions with N$_2$ for energies applicable for the EBIT experiment, $\sim$ 10 eV/u.  For systems of more astrophysical relevance, such as collisions with H and He at energies around 1 keV/u, Figures 3 and 4 show that dominant capture is into $n=11-14$ and $9-11$, respectively.
\linebreak
\indent 
These dominant capture results are in good agreement with the \cite{war2} EBIT study.  For instance, \cite{war2} states that classical trajectory Monte Carlo (CTMC) calculations predict $n_{\textrm{max}}$ for single electron capture between Fe$^{26+}$ and He to be $n=9$, whereas their experiment observed $n=11$ and 12.  As shown in the MCLZ calculations in Figure 4, around $\sim$ 10 eV/u,  $n_{\textrm{max}}=11$ and 12; thus, MCLZ theoretical results are in excellent agreement with experiment.  Further, observed $n_{\textrm{max}}$ values from the EBIT study for Fe$^{26+}$ collisions with N$_2$ and H$_2$ are $n=14$.  MCLZ calculations again agree with experiment, as shown in Figures 2 and 5, respectively with $n_{\textrm{max}}$ predicted to be 13 or 14.  
\linebreak
\indent
CTMC calculations have been previously performed for several Fe$^{26+}$ charge exchange collisions in \citet{kats} and \citet{schultz}.  As shown in Figure 3, total CTMC SEC cross sections are in excellent agreement with MCLZ total cross sections for Fe$^{26+}$ collisions with H.  Similarly, in Figures 4 and 5, CTMC results are compared to MCLZ for Fe$^{26+}$ collisions with He and H$_2$, respectively.  For both of these collisions, MCLZ predicts somewhat larger total cross sections than does CTMC.  It is expected that the CTMC method will not be reliable for collision energies $\lesssim$ 10 keV/u due to
the neglect of tunneling effects, while MCLZ calculations do not perform well for incident energies $\gtrsim$ 5 keV/u \citep{wu-stancil}.  
\linebreak
\indent
An $n\ell$-resolved cross section analysis is performed in Figure 6 to compare the various $\ell$-distribution models.  It is worth noting that the distributions depicted in Figure 6 apply to all Fe$^{26+}$ charge exchange collisions with no restriction to collision energy or target.  This universality is a downfall of the distribution models as they are independent of any parameters corresponding to these important factors.  We see variance in each $\ell$-distribution model; therefore, it is difficult to suggest which model is recommended until applying these various distributions to the cascade model for X-ray emission and comparing theoretical spectra to experimental data.  Therefore, by applying the various $n\ell$-resolved cross sections to populate atomic levels, theoretical spectra are obtained and displayed in Figure 7 for Fe$^{26+}$ and N$_2$ charge exchange collisions.  In the same figure, the experimental spectrum for the system is given as extracted from the EBIT experiment \citep{1} which shows a strong peak near 9220 eV due
to the Ly$\mu$ ($13p \rightarrow 1s$) and Ly$\nu$ ($14p\rightarrow 1s$) transitions. The intensity of this line is nearly as strong as Ly$\alpha$ at 6958 eV. To simulate the experimental conditions, we
adopted the low-energy $\ell$-distribution function for MCLZ $n$-resolved cross sections at 10 eV/u which results in a line intensity at 9220 eV more than a factor of 2 smaller than the
experiment (normalizing to Ly$\alpha$ and assuming a full width half maximum (FWHM) line width of 150 eV). Similar results are obtained for 100 eV/u. Theoretical spectra using CTMC cross sections for
Fe$^{26+}$ + H show a similar behavior, but with an even smaller intensity at 9220 eV \citep{1}. To produce a strong line near 9220 eV requires large charge exchange cross sections to the 13$p$ and 14$p$,
but Figure 6 indicates that the low-energy distribution peaks at $\ell=2$. All other distributions peak at even larger values of $\ell$. We postulate that to get agreement with the EBIT high-$n$ emission peak
that the cross sections to $p$-states should be enhanced. As a numerical experiment, we constructed an ad hoc $\ell$-distribution function $W^{\rm Low}_{n\ell^\prime}$ in which
$\ell^\prime = \ell -1$. $W^{\rm Low}_{n,\ell=0}$ is added to $W^{\rm Low}_{n,\ell^\prime=1}$ to preserve the normalization and  $W^{\rm Low}_{n,\ell^\prime=n-1}=0$. We refer to this as the {\it shifted low-energy} (SL1) distribution. Applying SL1 in the cascade/X-ray model gives the SL1 spectrum at 10 eV/u which is shown in Figure 7 to be in better agreement with the EBIT result. SL1
does underestimate Ly$\beta$ and Ly$\gamma$, but further manipulation of the distribution function is not warranted given the uncertainties in the experimental spectra.
\linebreak
\indent
Rather than presenting the spectra for all of the collision systems studied, Table 1 gives line ratios for X-ray emission resulting from SEC collisions between Fe$^{26+}$ and H, He, H$_2$, N$_2$, H$_2$O, and CO at three representative collision energies 10, 100, and 1000~eV/u.  These line ratios were obtained by applying the low energy distribution (Equation 10) to MCLZ $n$-resolved cross sections and implementing the results into the X-ray cascade model. The line ratio depicts the ratio between a specific emission line and the $\rm{Ly} \alpha$ line.
\subsection{Fe XXV}
As was discussed for Fe XXVI, product ion states must first be determined for the He-like Fe$^{24+}$ which are produced following charge exchange with Fe$^{25+}$.  As determined by Wigner-Witmer rules, Fe$^{25+}$ collisions with H correlate to two molecular manifolds, $^1\Sigma^+$ and $^3\Sigma^+$, which connect via radial couplings to molecular states of the same symmetry of
the asymptotic product states of Fe$^{24+}$ + H$^+$. As spin-changing collisions are forbidden, singlet and triplet MCLZ calculations are performed separately with the singlet cross sections
weighted by an approach probability factor of $1/4$, while the triplets are weighted by a factor of $3/4$. In contrast, collisions with helium, molecular hydrogen, molecular nitrogen, water, and carbon monoxide correlate to doublet molecular states with both Fe$^{24+}$ singlet and triplet states considered in a single MCLZ calculation.  Because both singlets and triplets must be considered, quantum defect theory is applied to estimate energies for all missing states of Fe$^{24+}$.  All Fe$^{24+}$ singlets states and their corresponding energies, as calculated by our implementation of quantum defect theory, are shown in Figure 8 and compared to available NIST data as well as energies estimated by the Cloudy spectral synthesis code
\citep{fer13}, which also implements a version of the quantum defect method.  Including all possible outgoing channels quickly makes the computation time significantly longer.  
\linebreak
\indent 
In comparing the total cross sections for Fe XXV collisions for multiple targets, we again see the interesting trend that with increasing ionization potential of the target, the total cross section decreases for collision energies $\gtrsim$ 10 eV/u as shown in Figure 9.  

Cross sections at the $n$-resolved level for each collision system are also computed.  In Figure 10, we display the $n$-resolved cross sections for Fe$^{25+}$ collisions with N$_2$.  Around $\sim$ 10 eV/u, which we again anticipate being representative of EBIT conditions, we see that dominant capture occurs into $n= 12-14$.  Again, we illustrate that dominant capture occurs into states for which energies are not supplied by NIST.  Thus, we see that through the quantum defect method, we can rapidly estimate the energies of these high-lying Rydberg states that assume vital roles in the Fe$^{25+}$ charge exchange process.  For astrophysical conditions $\sim$ 1 keV/u, Figures 11 and 12 display $n$-resolved cross sections for Fe$^{25+}$ collisions with H and He, respectively.  Dominant capture for these processes occurs into $n=11-14$ and $9-11$, respectively.
\linebreak
\indent
Charge exchange collisions with H-like projectile ions result in non-degenerate product ions.  Thus, the use of $\ell$-distribution functions is not required nor recommended.  However, it is commonplace for astrophysical modelers to incorporate such distribution functions for all charge exchange collisions -- even those with multielectron product ions due to the general lack of such data \citep{7}.  Therefore, in Figure 13, we compare the results from $n \ell S$-resolved cross sections via MCLZ to $n \ell$-resolved distribution functions for Fe$^{25+}$ collisions with N$_2$.  In this figure, the three dominant $n$-levels for the collision are shown at three different energies. After comparing the distributions, we see that the low energy distribution function compares best to the explicit MCLZ calculations.  When performing similar analyses to the distributions resulting from Fe$^{25+}$ CX collisions with H, He, H$_2$, H$_2$O, and CO, some of the distributions tend towards the low energy modification function or the separable function.   However, again, although it is not recommended, if one \textit{does} use a distribution function for non-bare projectile ion charge exchange collisions, we suggest using the low energy distribution function, as given by Equation (10), in most cases.  
\linebreak
\indent
Using the cascade model and $n \ell S$-resolved cross sections, we obtain theoretical X-ray spectra as given by Figure 14 for Fe$^{25+}$ collisions with N$_2$ with a FWHM of 150 eV.  These theoretical spectra are shown alongside the \cite{1} Fe XXV EBIT experiment.  We see reasonable agreement between theoretical and experimental spectra; however, all theoretical models overestimate the experimental intensity of the K$\beta$ and higher lines with the
overestimation increasing slightly with collision energy. As the K$\beta$ and higher lines originate only from Fe$^{24+}$ singlet states, the comparison suggests that the MCLZ calculation
predicts cross sections to $n~^1P$ with $n\sim12-14$ that are somewhat too large. 
\linebreak 
\indent 
Again, we include line ratios for the various Fe$^{25+}$ collision systems. MCLZ $n \ell S$-resolved cross sections were applied in the X-ray cascade model to obtain these ratios.  Table 2 gives the line ratios for X-ray emission resulting from SEC collisions between Fe$^{25+}$ and H, He, H$_2$, N$_2$, H$_2$O, and CO at the same representative collision energies used in Table 1.  Here, the line ratios depict the ratio between a specific emission line and the $\rm{K} \alpha$ resonant line. 
\linebreak
\indent 
Another useful diagnostic is the \textit{G}-ratio which is the ratio between the sum of the forbidden and intercombination line intensities to that of the resonant line \citep{grat, porquet, porter}.  Figure 15 displays the \textit{G}-ratios as a function of collision energy for emission resulting from the studied Fe$^{25+}$ single electron capture processes.  Also included in Figure 15 are the \textit{G}-ratio results presented in \cite{grat} pertaining to electron impact excitation (EIE, or so-called thermal excitation) for a plasma in 
collisional ionization equilibrium (CIE), as well as the \textit{G}-ratio for photoionization equilibrium (PIE) or a recombining photoionized plasma  \citep{bau00,porter}.  MCLZ charge exchange cross sections and their implementation into the X-ray cascade model yield similar $G\sim 0.8$ among He, H$_2$, N$_2$, H$_2$O, and CO targets with only a slight increase with
collision energy. These fall near the maximum EIE $G$-ratio which occurs for electron temperatures of $\sim$10$^7$~K \citep{grat}. Conversely, the \textit{G}-ratio for Fe$^{25+}$ and H charge exchange 
is  $\sim$3.7, similar to the values obtained for a recombining plasma (PIE) with an electron temperature of 10$^7$~K \citep{bau00,porter}.

Unfortunately, the resolution of the EBIT measurement was insufficient to resolve the individual K$\alpha$ components to determine a $G$-ratio. However, it is possible to
estimate the $G$-ratio from knowledge of the K$\alpha$ line centroid. For Fe$^{25+}$ collisions with N$_2$ at 10, 100, and 1000 eV/u, the centroids using the MCLZ cross
sections are predicted to be 6686.3, 6685.3, and 6684.0 eV, respectively.  These can be compared to the measured values for charge exchange,  $6666 \pm 5$ eV (EBIT beam off), 
and EIE, $6685 \pm 2.5$ eV  (EBIT beam on) \citep{1}. From
the experimental centroid, we estimate $G =1.2-2.65$, depending on the assumed relative forbidden and intercombination line intensities. This is somewhat
larger than the computed $G$-ratios from the MCLZ
calculations of 0.82, 0.92, and 1.05 for 10, 100, and 1000 eV/u, respectively.  
\section{Total Fe XXV and Fe XXVI Spectra}
The total theoretical spectrum, sum of Fe XXV and XXVI, is compared to the total EBIT experimental spectrum in Figure 16.  This total theoretical spectrum utilizes the shifted low energy (SL1) distribution as discussed for Fe$^{26+}$ collisions with N$_2$ and the $n \ell S$-resolved cross sections for Fe$^{25+}$ collisions with N$_2$.  By summing the two theoretical spectra and eliminating any need for extraction of the associated product ion's contribution to the emission, we see excellent agreement between theory and experiment.  Figure 17 compiles all theoretical total and individual spectra with corresponding emission lines and the total experimental spectra with associated Fe XXV and extracted Fe XXVI from \cite{1}.  From this figure, we can see each ion's contribution, theoretically and experimentally, to the overall spectra
and the very good agreement overall for this system of two highly-charged ions.

\section{Caveats}

A number of approximations were required in both the cross section calculations and the X-ray
spectra models. They are of such a magnitude that it is not meaningful to attempt to
estimate the theoretical uncertainty. Six main approximations are inherent in the current
MCLZ calculations: i) the splitting in the adiabatic potentials at the avoiding-crossing 
locations (Equation 4), ii) the adoption of the SEC MCLZ probability formalism of \citet{5},  iii)
for H-like product ions, the adoption of an $\ell$-distribution model, iv) for He-like
product ions, the estimation of missing Rydberg energies with the quantum defect approach, 
v) the neglect of multielectron capture (MEC) processes, and vi) the applicability of MCLZ calculations in the high energy regime. 
Of these, it is likely that item i) has the largest impact on all state-resolved cross sections.
For the case of the multi-channel probability model ii), only long-range avoided-crossings are
considered. The \citet{5} model cannot treat short-range avoided-crossings which are
usually highly endoergic. As a consequence, the cross sections for the highest $n$, those
which fall-off as $E^{-1/2}$ for all energies, will be lower-limits for the highest collision
energies. MEC processes v) also cannot be treated with the SEC MCLZ probability model.  There exists no analytic MEC formula valid for all ions -- each charge exchange collision system must be treated individually and therefore such calculations are not amendable to our streamlined program set.
However, \citet{1} argue that MEC processes do not play a significant role for the Fe$^{25+}$ and Fe$^{26+}$ systems. On the other hand, in a combined experimental/theoretical study of Ne$^{10+}$ collisions  with He,
Ne, and Ar, \cite{ali05} find that MEC could contribute from $\sim$10-50\% of the X-ray spectrum with the fraction increasing with the number of electrons in the target.  Finally, the performance of the multi-channel calculations in a molecular representation, such as the MCLZ method, was vi) suggested by \cite{wu-stancil} to break down for incident energies $\gtrsim$ 5 keV/u.  At such high incident energies, other channels (ionization, excitation, dissociation) become important and eventually dominant.  This leads to the interpretation that MCLZ CX cross sections are upper limits in the high energy regime.  MCLZ calculations are compared to CTMC data (more applicable at higher collision energies) in Figures 3, 4, \& 5 and show good agreement thus meriting their use as a base approximation for such high collision energies.

In the cascade/X-ray emission models, we have assumed a low-density plasma environment
in which only single collision events occur and  the plasma is optically thin. Optical
depth effects and/or multiple collisions may become important thus modifying
the predicted line ratios in high density environments. In addition, as detailed above, the cross sections were only computed for SEC so that no MEC effects are included in the X-ray spectra or line ratios.  

\section{Conclusions}
Fe$^{25+}$ and Fe$^{26+}$ charge exchange collisions with multiple targets were studied because of their astrophysical relevance in the Galactic center, the Galactic ridge, and other environments; the computational difficulty that such highly charged systems ensued; and the presence of experimental data for comparison.  Theoretical studies using quantum defect theory and the multi-channel Landau-Zener approach were able to compute, with ease, $n \ell S$-resolved cross sections for Fe$^{25+}$ and Fe$^{26+}$ collisions with H, He, H$_2$, N$_2$, H$_2$O, and CO.  This is the first time, to our knowledge, that MCLZ calculations have been performed for these systems.  Further, to our knowledge, this is the first time that such calculations have been performed altogether for Fe$^{26+}$ collisions with N$_2$, H$_2$O, and CO and Fe$^{25+}$ collisions with all targets considered in this work.
\linebreak
With this new data and approach, comparisons of total Fe XXV and Fe XXVI theoretical spectra to total experimental EBIT spectra demonstrate very good agreement.  Through the simplicity of the MCLZ method, this data was generated quickly and robustly thus allowing interesting trends to be identified.  Namely, with increasing ionization potential of the target, we see decreasing total cross sections for collision energies $\gtrsim$ 10 eV/u.  Further,  several $\ell$-distribution models have been tested for Fe$^{25+}$ and Fe$^{26+}$ single electron capture.  Such analysis suggests that commonly used $\ell$-distribution models struggle to accurately reflect the true distribution of electron capture as understood by more advanced frameworks of theory. However, if there is no other recourse, then the so-called low-energy
distribution function is preferred. 
\linebreak
Ultimately, we have shown through extensive theoretical studies, agreement between theory and experiment, and the complexity of the Fe XXV and Fe XXVI systems, that the quantum defect/ multi-channel Landau-Zener approach is an excellent tool to quickly provide accurate charge exchange data, for any system, to experimentalists, astrophysical modelers, and astronomers and aid in bringing resolution to the current lack of atomic and molecular data for such a vital process.  

\acknowledgments
This work was partially supported by NASA grants NNX09AC46G and NNX13AF31G.
We thank Brad Wargelin for providing the experimental EBIT spectra and for
helpful discussions.

\newpage

\begin{figure}[h!]
\centering
\includegraphics[scale=0.4]{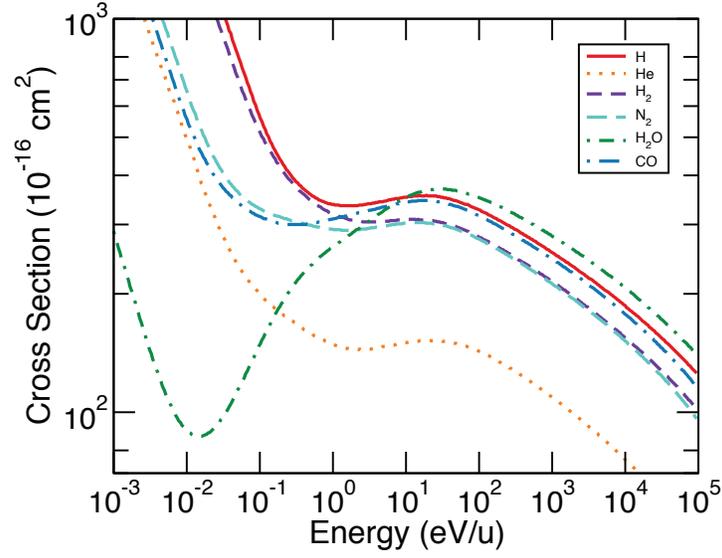}
\caption{Total SEC MCLZ cross sections for Fe$^{26+}$ collisions with H, He, H$_2$, N$_2$, H$_2$O, and CO.}
\label{fig1}
\end{figure}

\begin{figure}[h!]
\centering
\includegraphics[scale=0.4]{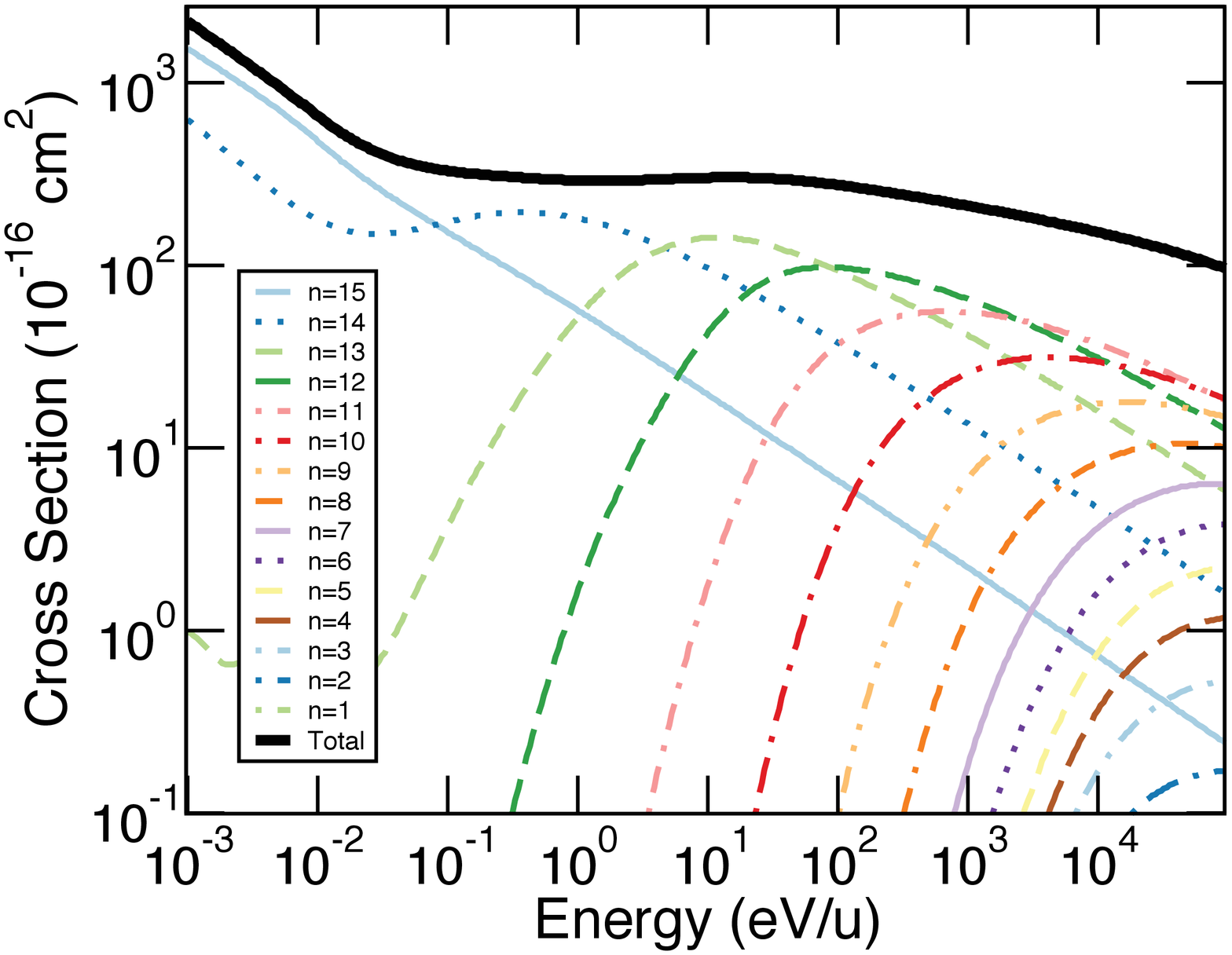}
\caption{SEC $n$-resolved MCLZ cross sections for Fe$^{26+}$ collisions with N$_2$.}
\label{fig2}
\end{figure}

\begin{figure}[h!]
\centering
\includegraphics[scale=0.4]{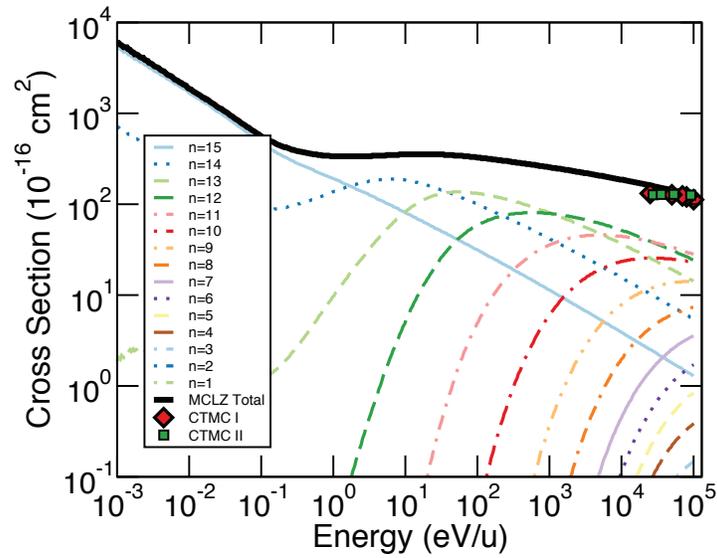}
\caption{SEC $n$-resolved MCLZ cross sections for Fe$^{26+}$ collisions with H.  For reference, CTMC SEC total cross sections, as calculated in \cite{kats} (red diamonds -- CTMC I) and \cite{schultz} (green squares -- CTMC II), are given. See text for details.}
\label{fig3}
\end{figure}

\begin{figure}[h!]
\centering
\includegraphics[scale=0.4]{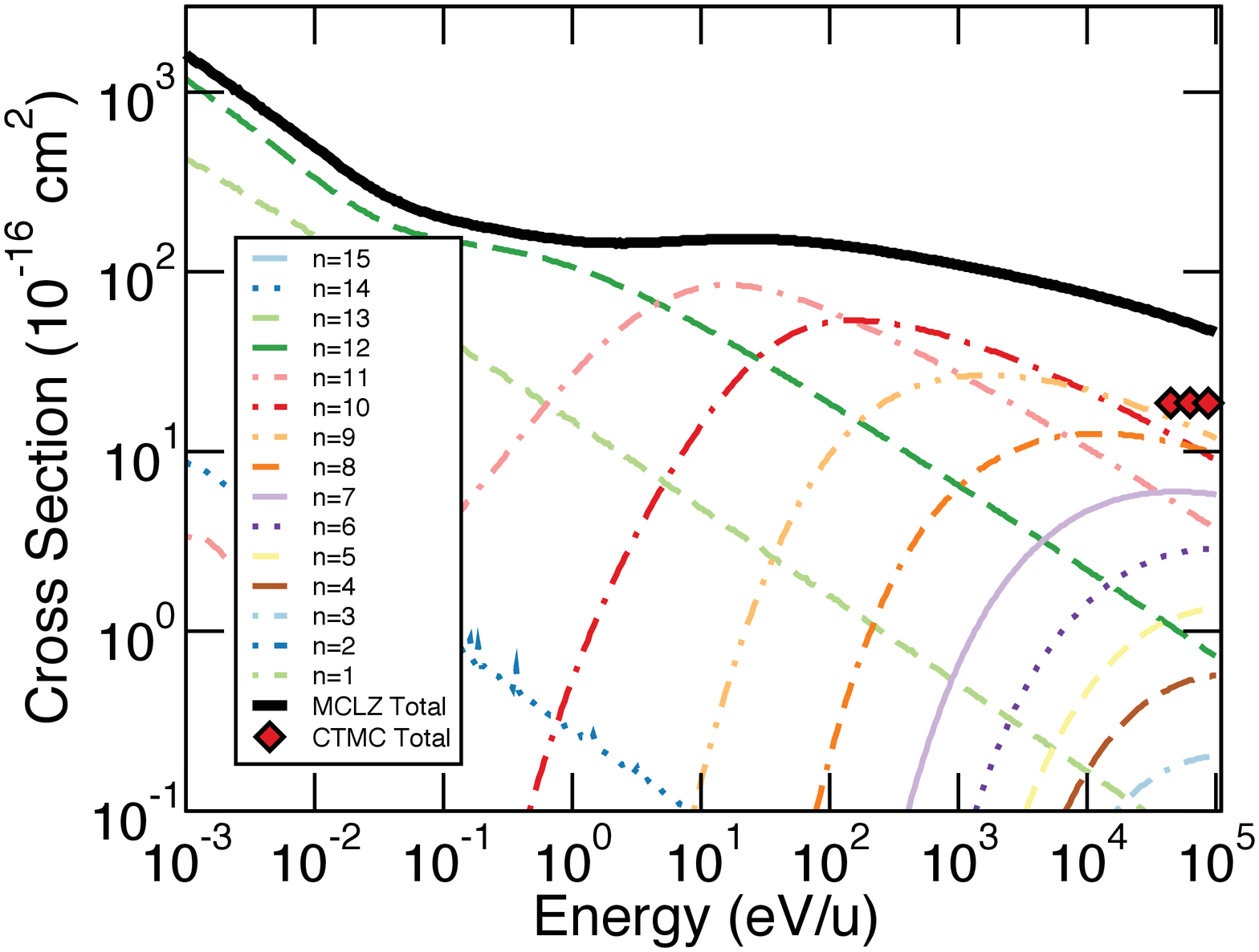}
\caption{SEC $n$-resolved MCLZ cross sections for Fe$^{26+}$ collisions with He. For reference, CTMC SEC total cross sections, as calculated in \cite{schultz} (red diamonds) are given.  See text for details.}
\label{fig4}
\end{figure}

\begin{figure}[h!]
\centering
\includegraphics[scale=0.4]{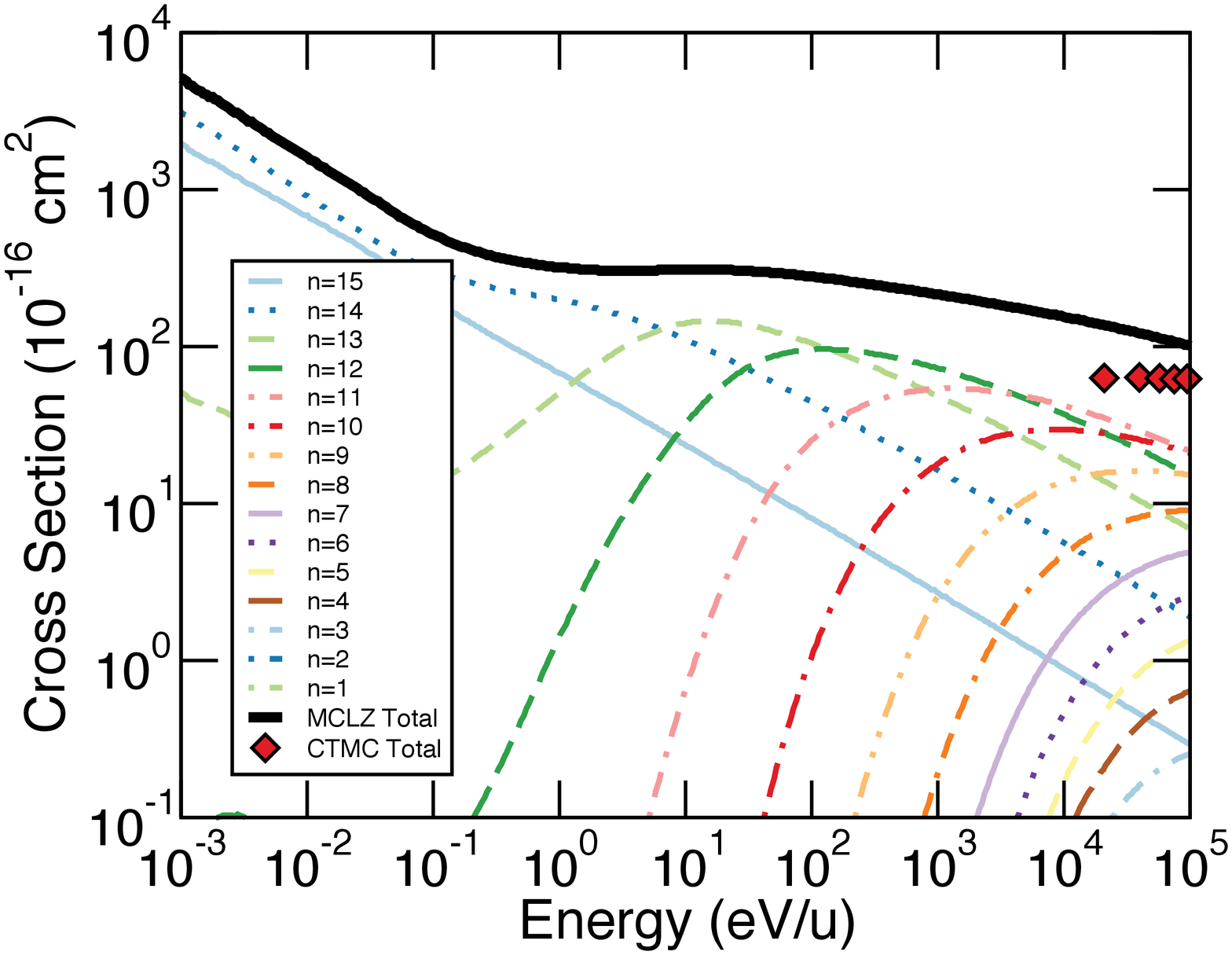}
\caption{SEC $n$-resolved MCLZ cross sections for Fe$^{26+}$ collisions with H$_2$. For reference, CTMC SEC total cross sections, as calculated in \cite{schultz} (red diamonds) are given.  See text for details.}
\label{fig5}
\end{figure}

\begin{figure}[h!]
\centering
\includegraphics[scale=0.7]{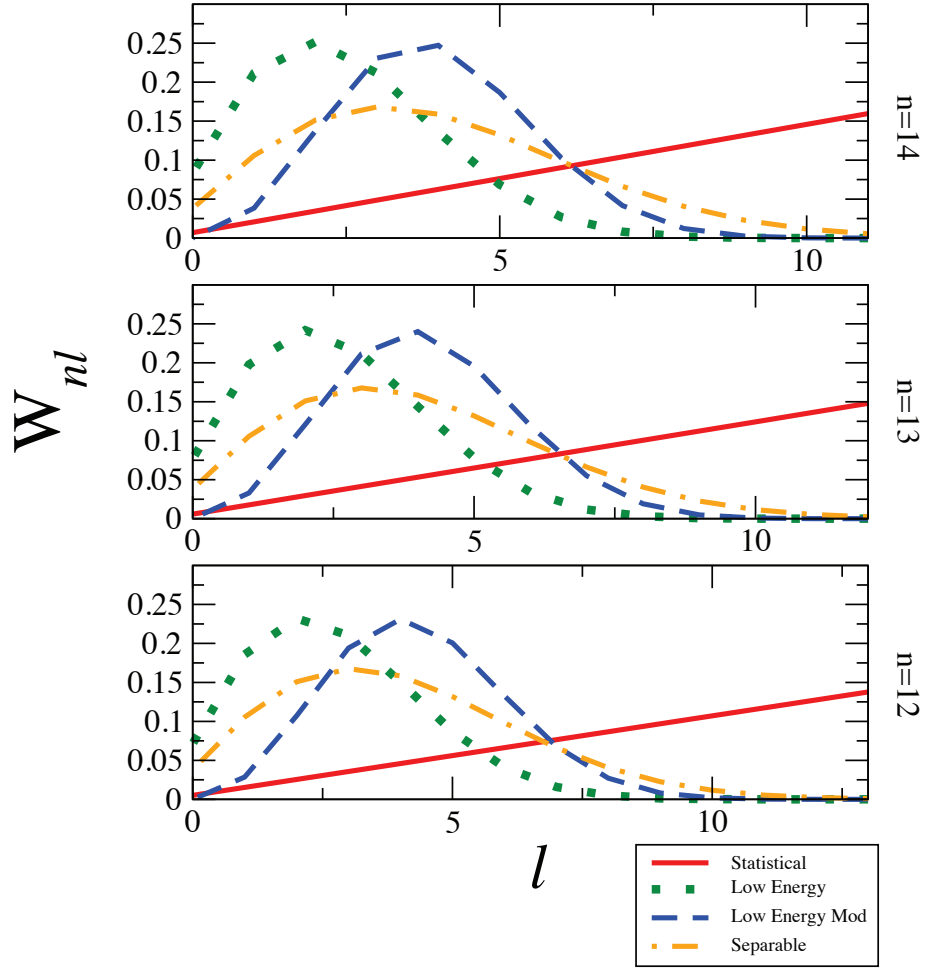}
\caption{Analytical $\ell$-distribution functions $W_{n\ell}$ for bare-ion collisions for $q_1$ = 26.}
\label{fig8}
\end{figure}

\begin{figure}[h!]
\centering
\includegraphics[scale=0.6]{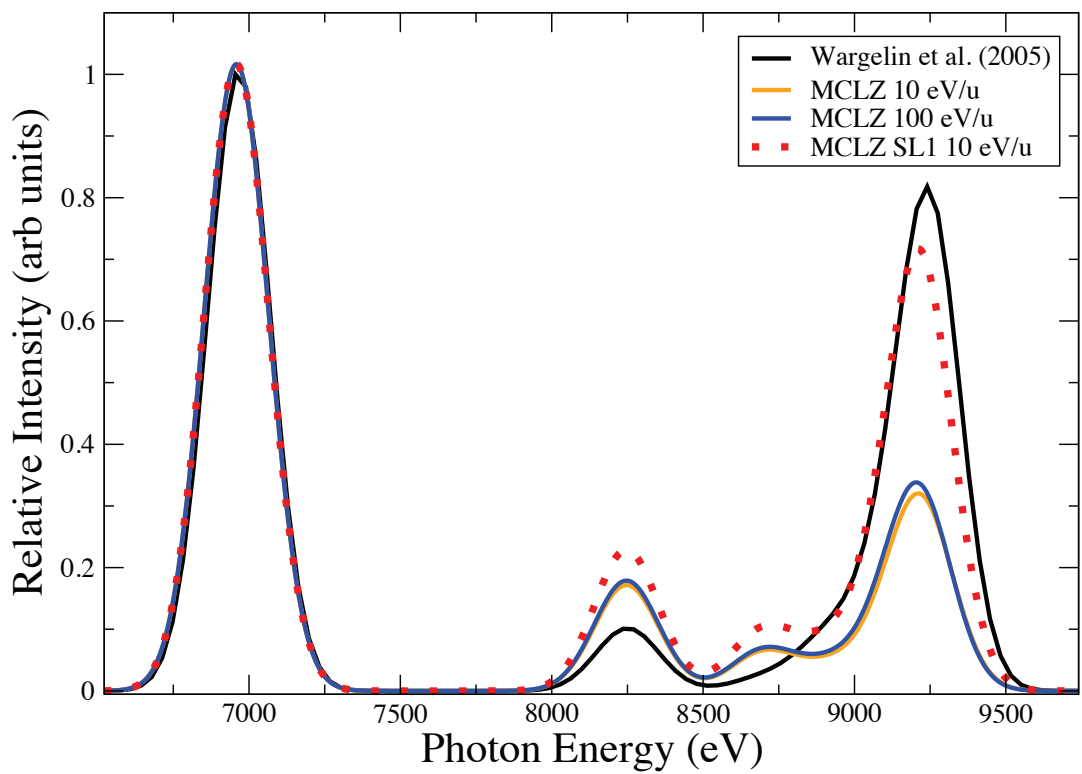}
\caption{Comparison of theoretical MCLZ/cascade model spectrum to \cite{1} experimental spectrum for Fe$^{26+}$ collisions with N$_2$.  All spectra
are normalized to Ly$\alpha$ and the MCLZ cross sections used the low-energy $\ell$-distribution or
shifted low-energy (SL1) distribution functions. A FWHM of 150 eV was assumed for the theoretical spectra. See text for details.}
\label{fig9}
\end{figure}

\begin{figure}[h!]
\centering
\includegraphics[scale=0.6]{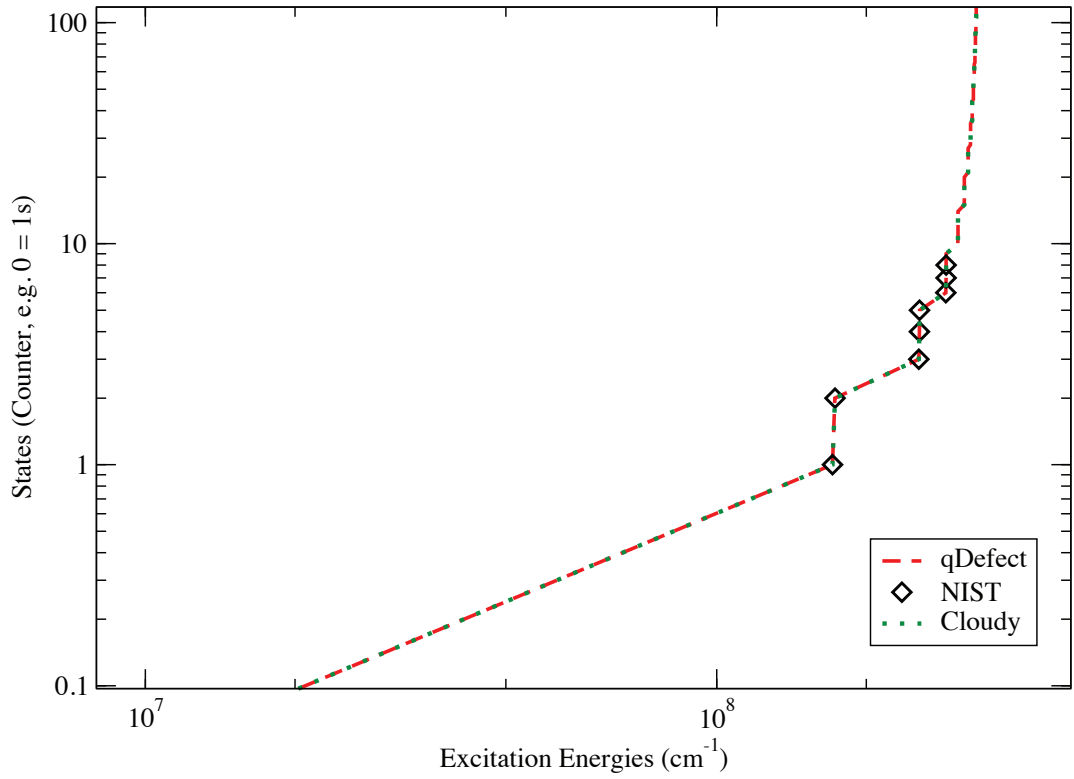}
\caption{qDefect excitation energies of He-like Fe$^{24+}$ singlets as compared to NIST data and quantum defect approximations from the Cloudy spectral synthesis code.}
\label{fig10}
\end{figure}

\begin{figure}[h!]
\centering
\includegraphics[scale=0.4]{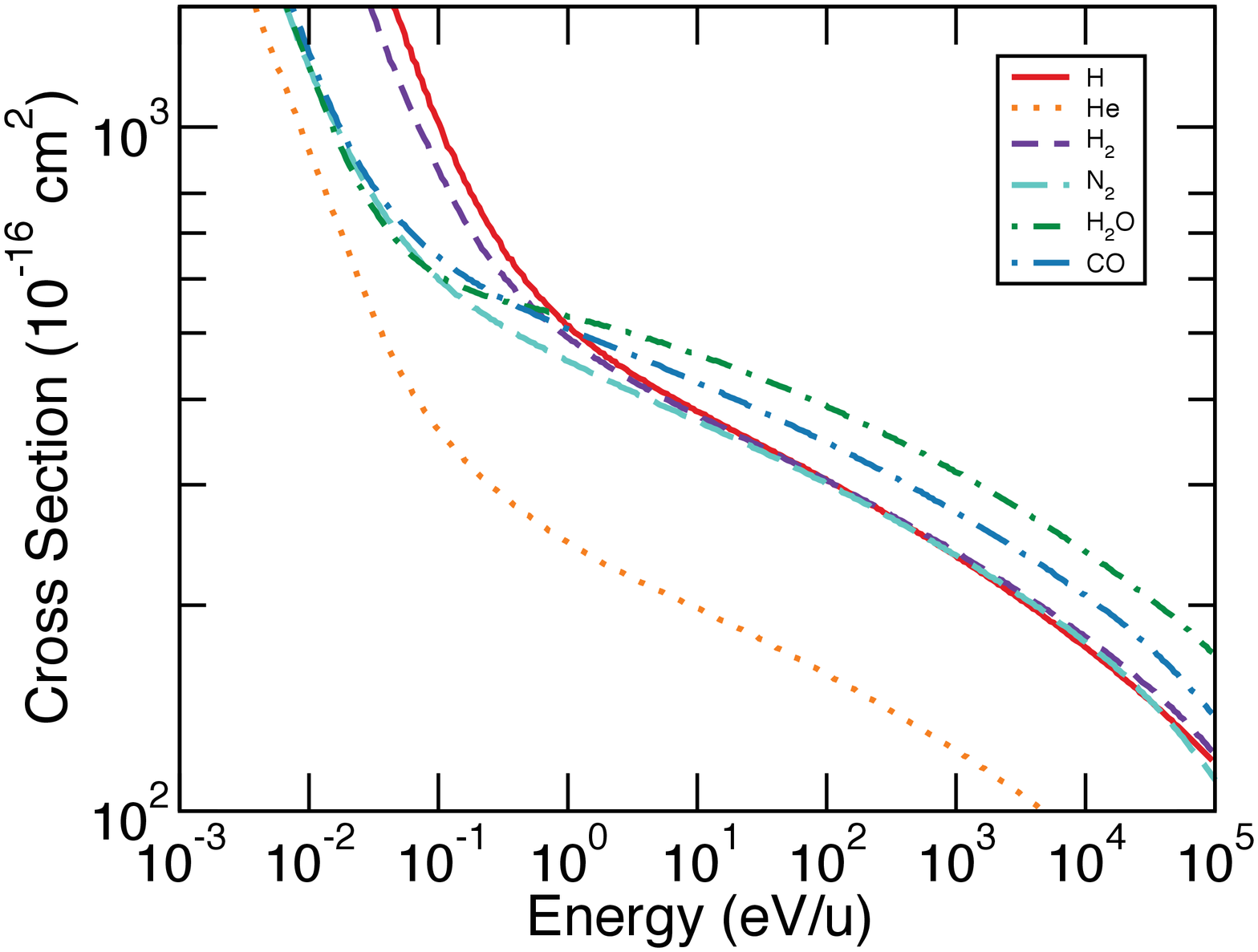}
\caption{Total SEC MCLZ cross sections for Fe$^{25+}$ collisions with H, He, H$_2$, N$_2$, H$_2$O, and CO.}
\label{fig11}
\end{figure}

\begin{figure}[h!]
\centering
\includegraphics[scale=0.4]{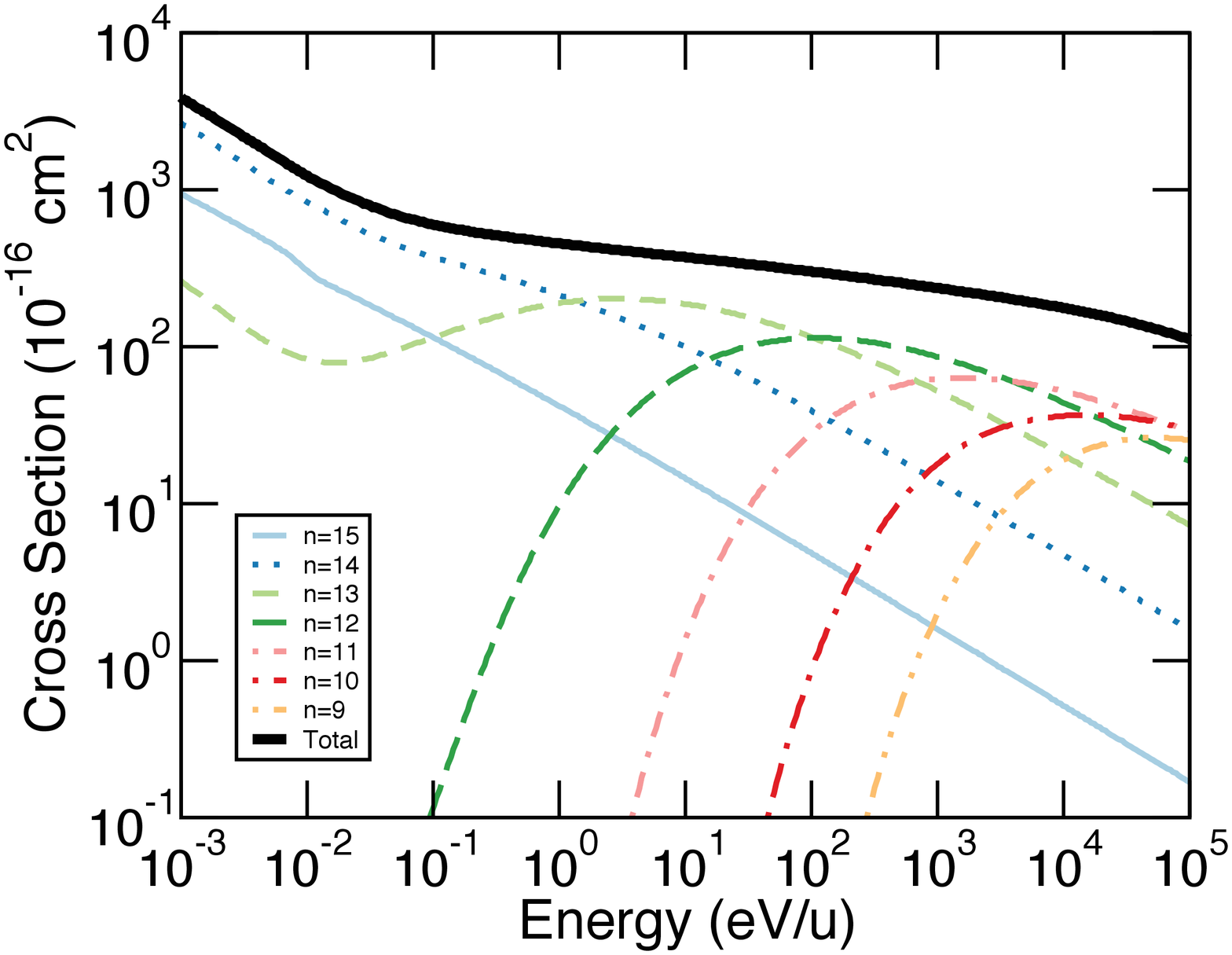}
\caption{SEC $n$-resolved MCLZ cross sections for Fe$^{25+}$ collisions with N$_2$.}
\label{fig12}
\end{figure}

\clearpage

\begin{figure}[h!]
\centering
\includegraphics[scale=0.4]{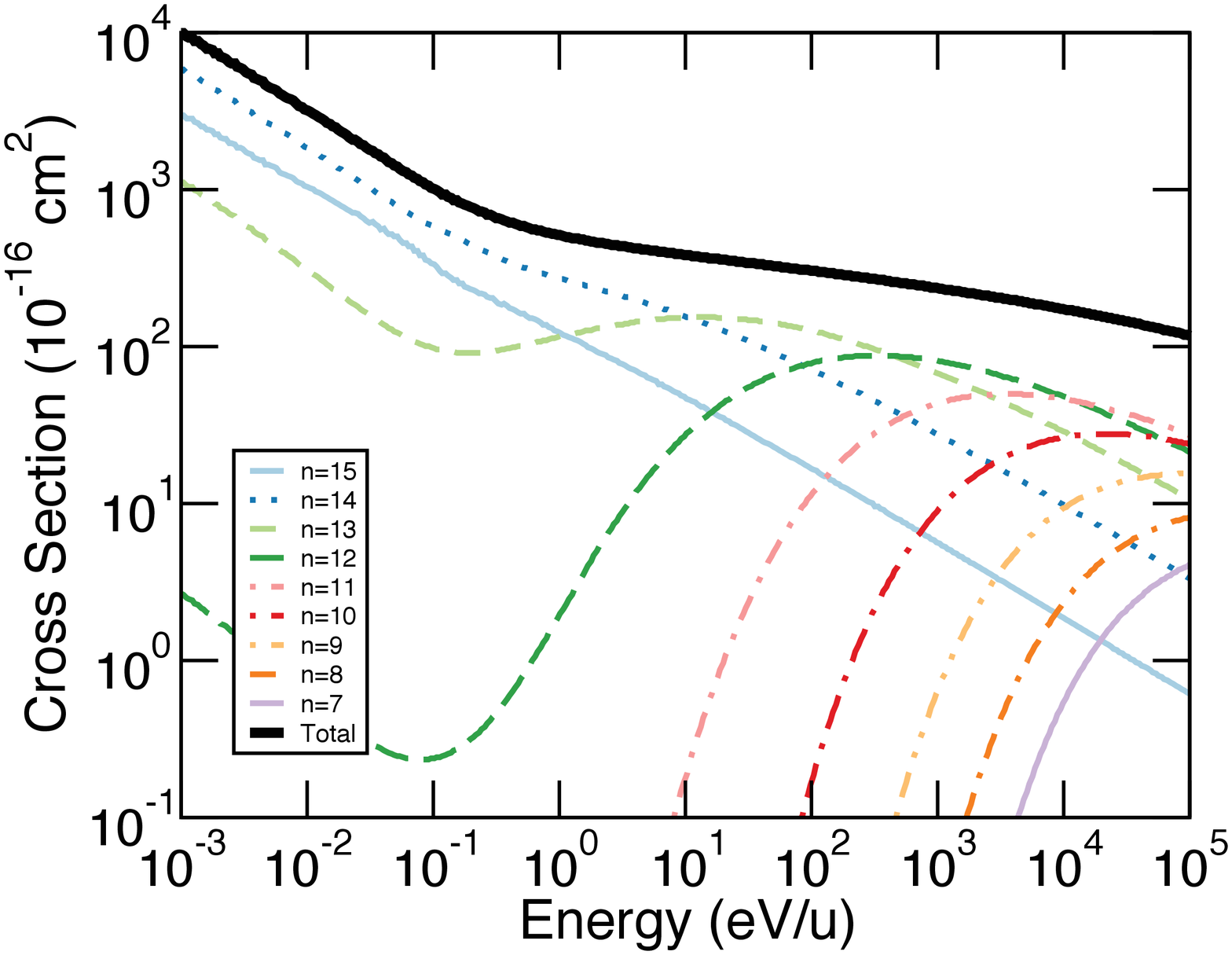}
\caption{SEC $n$-resolved MCLZ cross sections for Fe$^{25+}$ collisions with H.}
\label{fig16}
\end{figure}

\begin{figure}[h!]
\centering
\includegraphics[scale=0.4]{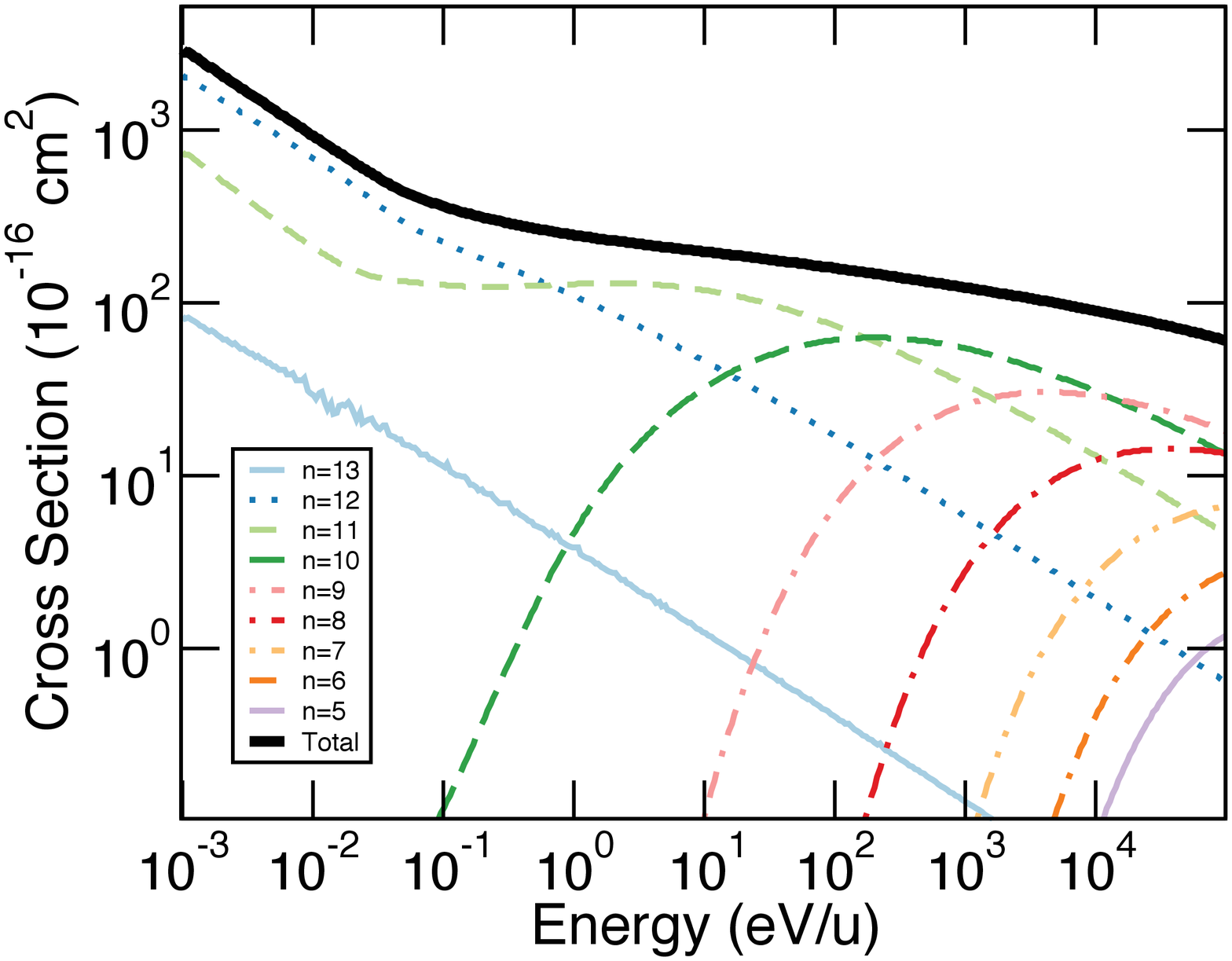}
\caption{SEC $n$-resolved MCLZ cross sections for Fe$^{25+}$ collisions with He.}
\label{fig17}
\end{figure}

\begin{figure}[h!]
\centering
\includegraphics[scale=0.6]{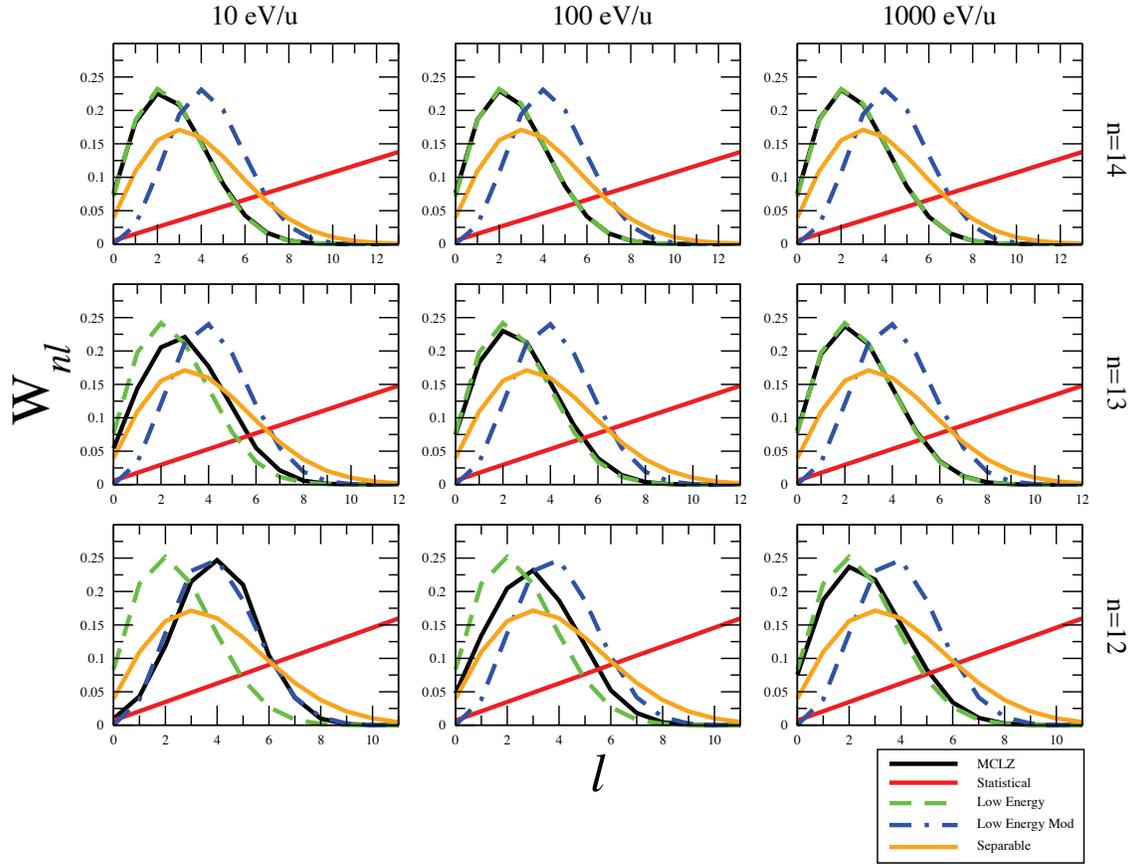}
\caption{Comparison of $n \ell$-resolved MCLZ cross sections with analytical $\ell$-distribution functions for Fe$^{25+}$ SEC collisions with N$_2$.}
\label{fig21}
\end{figure}

\begin{figure}[h!]
\centering
\includegraphics[scale=0.6]{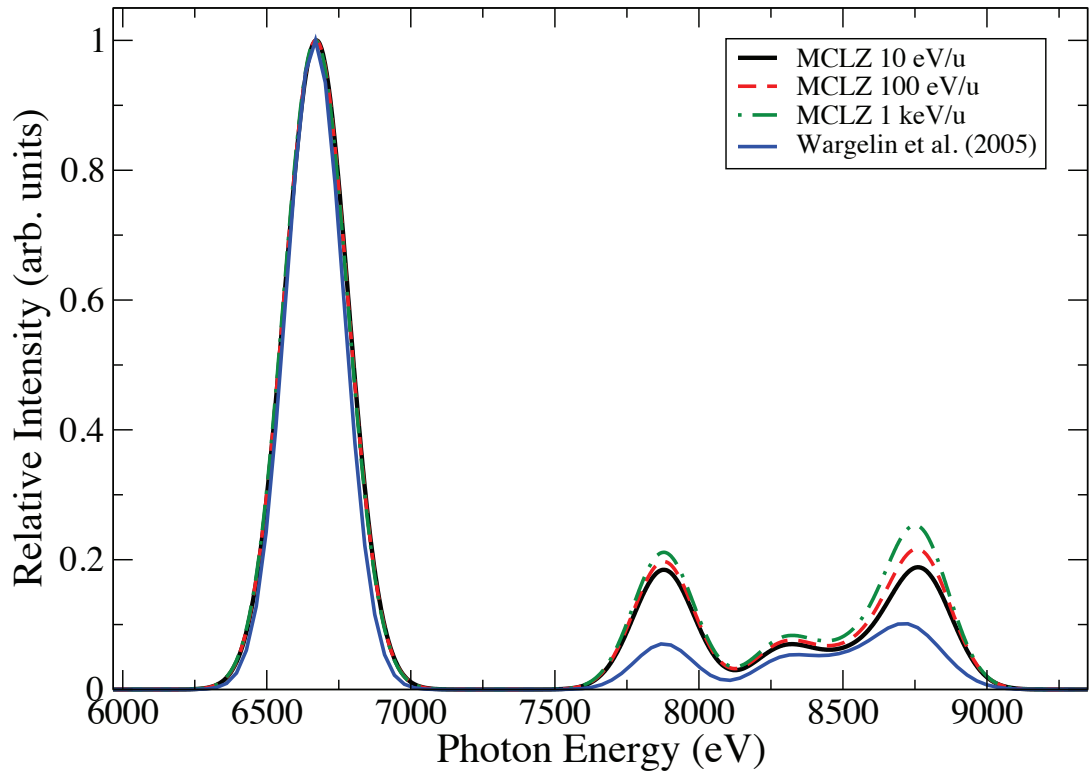}
\caption{Comparison of theoretical MCLZ/cascade model spectrum to \cite{1} experimental spectrum for Fe$^{25+}$ SEC collisions with N$_2$.  All spectra
are normalized to K$\alpha$.  A FWHM of 150 eV was assumed for the theoretical spectra. See text for details.}
\label{fig24}
\end{figure}

\begin{figure}[h!]
\centering
\includegraphics[scale=0.6]{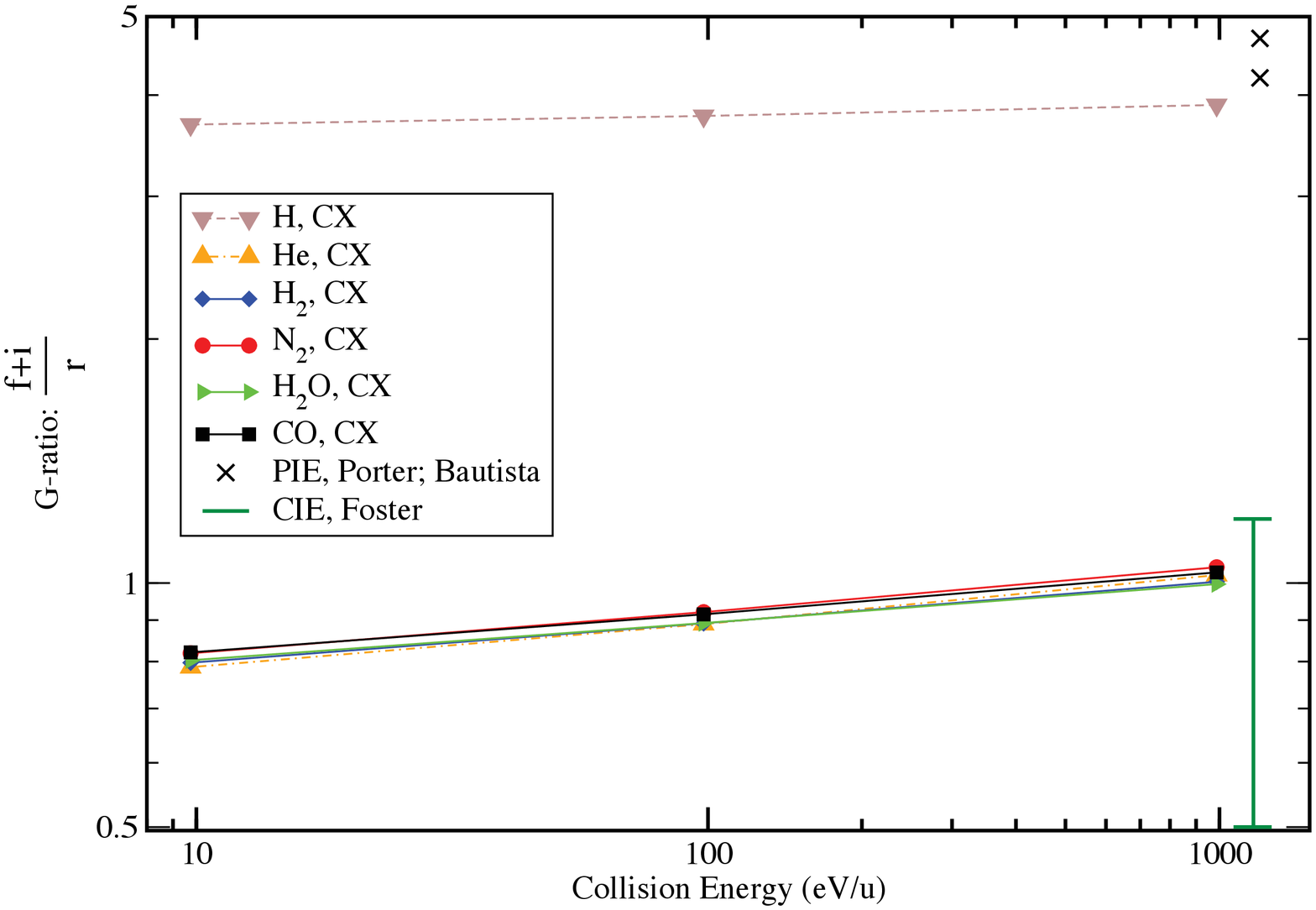}
\caption{\textit{G}-ratios calculated for Fe$^{25+}$ SEC charge exchange collisions with several targets.  Note that \textit{G}-ratios for He, H$_2$, N$_2$, H$_2$O, and CO are nearly overlapping.  For reference,  \textit{G}-ratios for collisional ionization equilibrium (CIE) by EIE \citep{grat} and photoionization equilibrium (PIE) for a recombining plasma \citep{bau00,porter}, both for
electron temperatures of 10$^7$ K, are shown on the right.}
\label{fig25}
\end{figure}

\begin{figure}[h!]
\centering
\includegraphics[scale=0.6]{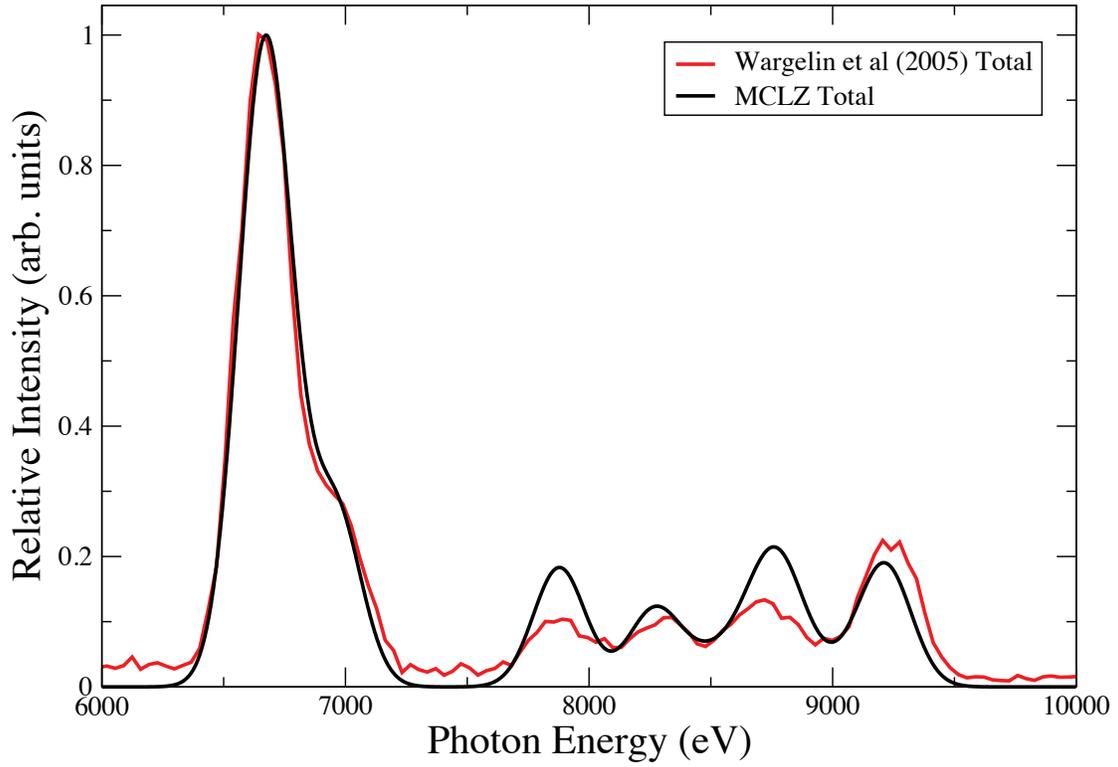}
\caption{Comparison of combined Fe XXV and Fe XXVI theoretical spectra resulting from SEC charge exchange collisions with N$_2$ to total experimental spectra from \cite{1}.
See text for details.}
\label{fig26}
\end{figure}

\begin{figure}[h!]
\centering
\includegraphics[scale=0.6]{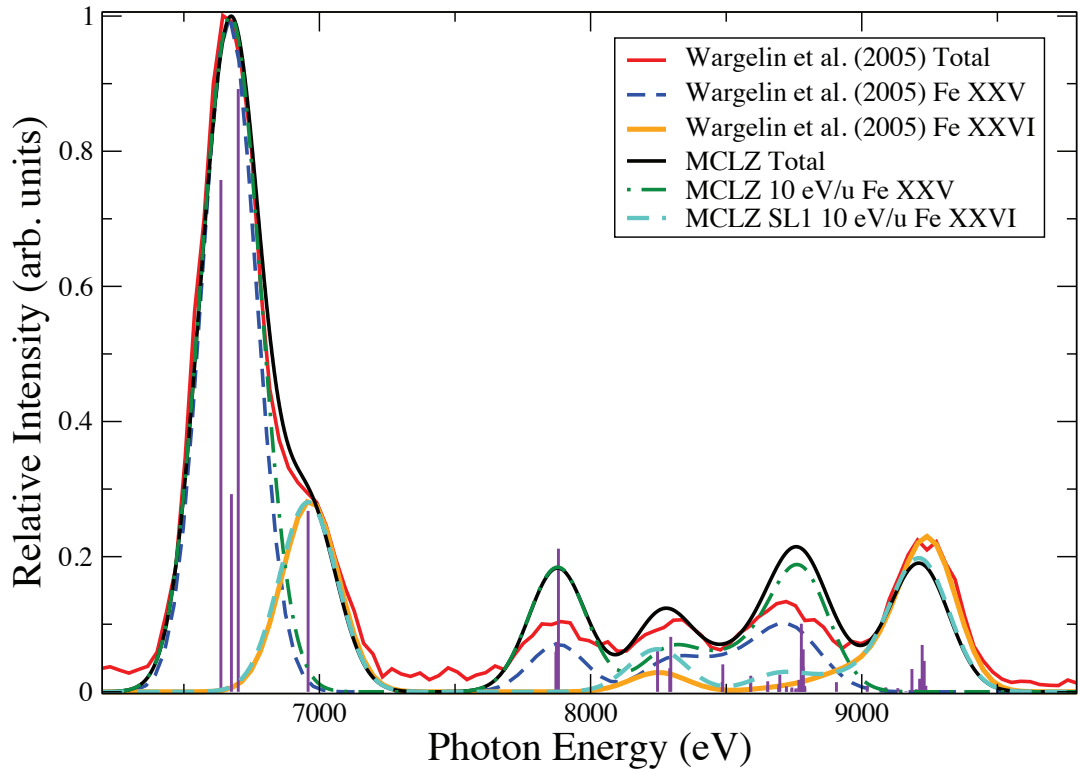}
\caption{Breakdown of total theoretical MCLZ SEC spectra into Fe XXV and Fe XXVI components with corresponding emission lines and total experimental spectra into the Fe XXVI extraction and Fe XXV from \cite{1}.}
\label{fig27}
\end{figure}

\clearpage

\begin{deluxetable}{c c c c c c}
\tabletypesize{\scriptsize}

\tablecaption{Charge Exchange X-ray emission normalized line ratios for Fe $\uppercase\expandafter{\romannumeral 26}$}
\tablewidth{0pt}
\tablehead{
\multicolumn{1}{c}{Photon Energy} &
\multicolumn{1}{c}{Neutral} & 
\multicolumn{1}{c}{Line}&
\multicolumn{1}{c}{}&
\multicolumn{1}{c}{Intensity}&
\multicolumn{1}{c}{}\\
\cline{4-6}
\multicolumn{1}{c}{(eV)} & 
\multicolumn{1}{c}{} & 
\multicolumn{1}{c}{} &
\multicolumn{1}{c}{10 eV/u} &
\multicolumn{1}{c}{100 eV/u} &
\multicolumn{1}{c}{1 keV/u}\\
\multicolumn{1}{c}{} & 
\multicolumn{1}{c}{} & 
\multicolumn{1}{c}{} &
\multicolumn{1}{c}{(43.8 km/s)} &
\multicolumn{1}{c}{(138 km/s)} &
\multicolumn{1}{c}{(438 km/s)}
}
\startdata
6958	&	H	&	 Ly$\alpha$       &	   1.0000	  &	1.0000	  &	1.0000	\\
8247	&	H	&	 Ly$\beta$	       & 	0.1646	  &	0.1693	  &	0.1756	\\
8698	&	H	&	 Ly$\gamma$   	& 	0.0289	  &	0.0607	  &	0.0642	\\
8907	&	H	&	 Ly$\delta$        &	   0.0172	  &	0.0304  &	0.0325	\\
9020	&	H	&	 Ly$\epsilon$	    &    0.0114	  &	0.0181	  &	0.0196	\\
9088	&	H	&	 Ly$\zeta$   	    & 	0.0583	  &	0.0121	  &	0.0132	\\
9133	&	H	&	 Ly$\eta$	        &	0.0090	  &	0.0094	  &	0.0099	\\
9163	&	H	&	 Ly$\theta$	        &	0.0017	  &	0.0025	  &	0.0062	\\
9185	&	H	&	 Ly$\iota$	        &	0.0413	  &	0.0444	  &	0.0442	\\
9201	&	H	&	 Ly$\kappa$	    &	   0.0050	  &	0.0086	  &	0.0460	\\
9213	&	H	&	 Ly$\lambda$	    &   0.0051	  &	0.0501	  &	0.0875	\\
9223	&	H	&	 Ly$\mu$	        &	0.0569	  &	0.1006	  &	0.0806	\\
9230	&	H	&	 Ly$\nu$	            &	0.1217	  &	0.0728	  &	0.0398	\\
9236	&	H	&	 Ly$\xi$	            &	0.0514	  &	0.0219	  &	0.0104	\\
\hline
6958	&	He	&	 Ly$\alpha$       &	   1.0000	  &	1.0000	  &	1.0000	\\
8247	&	He	&	 Ly$\beta$	       & 	0.1884	  &	0.1955	  &	0.2030	\\
8698	&	He	&	 Ly$\gamma$   	& 	0.0714  &	0.0756	  &	0.0801	\\
8907	&	He	&	 Ly$\delta$        &	   0.0370	  &	0.0397  &	0.0427	\\
9020	&	He	&	 Ly$\epsilon$	    &    0.0227  &	0.0247	  &	0.0270	\\
9088	&	He	&	 Ly$\zeta$   	    & 	0.0155  &	0.0170	  &	0.0213	\\
9133	&	He	&	 Ly$\eta$	        &	0.0108	  &	0.0128	  &	0.0372	\\
9163	&	He	&	 Ly$\theta$	        &	0.0153	  &	0.0388	  &	0.1026	\\
9185	&	He	&	 Ly$\iota$	        &	0.0439  &	0.1335	  &	0.1410	\\
9201	&	He	&	 Ly$\kappa$	    &	   0.1721	  &	0.1378	  &	0.0826	\\
9213	&	He	&	 Ly$\lambda$	    &   0.0973	  &	0.0387	  &	0.0181	\\
9223	&	He	&	 Ly$\mu$	        &   0.0085	  &	0.0030	  &	0.0013	\\
9230	&	He	&	 Ly$\nu$	            &   0.0002  &	5.2E-5	  &	2.2E-5	\\
9236	&	He	&	 Ly$\xi$	            &   2.5E-7	  &	8.5E-8	  &	3.6E-8	\\
\hline
6958	&	H$_2$	&	 Ly$\alpha$       &	   1.0000	  &	1.0000	  &	1.0000	\\
8247	&	H$_2$	&	 Ly$\beta$	       & 	0.1680	  &	0.1744  &	0.1816	\\
8698	&	H$_2$	&	 Ly$\gamma$   	& 	0.0600  &	0.0635	  &	0.0676	\\
8907	&	H$_2$	&	 Ly$\delta$        &	   0.0299	  &	0.0321  &	0.0346	\\
9020	&	H$_2$	&	 Ly$\epsilon$	    &    0.0178  &	0.0193	  &	0.0211	\\
9088	&	H$_2$	&	 Ly$\zeta$   	    & 	0.0118  &	0.0129	  &	0.0143	\\
9133	&	H$_2$	&	 Ly$\eta$	        &	0.0093	  &	0.0098	  &	0.0108	\\
9163	&	H$_2$	&	 Ly$\theta$	        &	0.0019	  &	0.0047	  &	0.0129	\\
9185	&	H$_2$	&	 Ly$\iota$	        &	0.0473  &	0.0415	  &	0.0558	\\
9201	&	H$_2$	&	 Ly$\kappa$	    &	   0.0044	  &	0.0309	  &	0.0804	\\
9213	&	H$_2$	&	 Ly$\lambda$	    &   0.0292	  &	0.0963	  &	0.0971	\\
9223	&	H$_2$	&	 Ly$\mu$	        &   0.1137	  &	0.0954	  &	0.0586	\\
9230	&	H$_2$	&	 Ly$\nu$	            &   0.0855  &	0.0385  &	0.0186\\
9236	&	H$_2$	&	 Ly$\xi$	            &   0.0173	  &	0.0066  &	0.0029\\
\hline
6958	&	N$_2$	&	 Ly$\alpha$ -- Low      &	   1.0000	  &	1.0000	  &	1.0000	\\
8247	&	N$_2$	&	 Ly$\beta$	-- Low        & 	0.1688	  &	0.1760  &	0.1840	\\
8698	&	N$_2$	&	 Ly$\gamma$ -- Low  	& 	0.0604  &	0.0644	  &	0.0690	\\
8907	&	N$_2$	&	 Ly$\delta$  -- Low      &	   0.0302	  &	0.0326  &	0.0355	\\
9020	&	N$_2$	&	 Ly$\epsilon$	-- Low    &    0.0180  &	0.0197  &	0.0218	\\
9088	&	N$_2$	&	 Ly$\zeta$ -- Low  	    & 	0.0119  &	0.0132	  &	0.0151	\\
9133	&	N$_2$	&	 Ly$\eta$	-- Low        & 	0.0093	  &	0.0099	  &	0.0129\\
9163	&	N$_2$	&	 Ly$\theta$	  -- Low      & 	0.0021	  &	0.0058	  &	0.0207	\\
9185	&	N$_2$	&	 Ly$\iota$	-- Low        &	    0.0471  &	0.0418  &	0.0657	\\
9201	&	N$_2$	&	 Ly$\kappa$	-- Low    &	   0.0054	  &	0.0439	  &	0.0842	\\
9213	&	N$_2$	&	 Ly$\lambda$	-- Low    &   0.0390	  &	0.0996	  &	0.0890	\\
9223	&	N$_2$	&	 Ly$\mu$	  -- Low      &   0.1171	  &	0.0866  &	0.0511	\\
9230	&	N$_2$	&	 Ly$\nu$	  -- Low          &   0.0763  &	0.0333  &	0.0159\\
9236	&	N$_2$	&	 Ly$\xi$	  -- Low          &   0.0149	  &	0.0056  &	0.0025\\
\hline 
6958	&	N$_2$	&	 Ly$\alpha$ -- SL1 	&	1.0000	&	1.0000	&	1.0000	\\
8247	&	N$_2$	&	 Ly$\beta$ -- SL1	&	0.2248	&	0.2371	&	0.2515	\\
8698	&	N$_2$	&	 Ly$\gamma$ -- SL1	&	0.0955	&	0.1037	&	0.1136	\\
8907	&	N$_2$	&	 Ly$\delta$	 -- SL1&	0.0532	&	0.0589	&	0.0660	\\
9020	&	N$_2$	&	 Ly$\epsilon$	 -- SL1&	0.0343	&	0.0385	&	0.0439	\\
9088	&	N$_2$	&	 Ly$\zeta$ -- SL1	&	0.0241	&	0.0275	&	0.0324	\\
9133	&	N$_2$	&	 Ly$\eta$ -- SL1	&	0.0200	&	0.0216	&	0.0285\\
9163	&	N$_2$	&	 Ly$\theta$	 -- SL1&	0.0038	&	0.01365&	0.0465	\\
9185	&	N$_2$	&	 Ly$\iota$	-- SL1&	0.1262	&	0.1090	&	0.1529	\\
9201	&	N$_2$	&	 Ly$\kappa$	-- SL1&	0.0140	&	0.0976	&	0.1852	\\
9213	&	N$_2$	&	 Ly$\lambda$	 -- SL1&	0.0723	&	0.1836	&	0.1650	\\
9223	&	N$_2$	&	 Ly$\mu$-- SL1	&	0.2588	&	0.1933	&	0.1150\\
9230	&	N$_2$	&	 Ly$\nu$-- SL1	&	0.1709	&	0.0754	&	0.0364	\\
9236	&	N$_2$	&	 Ly$\xi$-- SL1	&	0.0337	&	0.0128	&	0.0057	\\
\hline
6958	&	H$_2$O	&	 Ly$\alpha$ 	&	1.0000	&	1.0000	&	1.0000	\\
8247	&	H$_2$O	&	 Ly$\beta$	&	0.1634	&	0.1674	&	0.1735	\\
8698	&	H$_2$O	&	 Ly$\gamma$	&	0.0577	&	0.0597	&	0.0631	\\
8907	&	H$_2$O	&	 Ly$\delta$	&	0.0286	&	0.0298	&	0.0318	\\
9020	&	H$_2$O	&	 Ly$\epsilon$	&	0.0170	&	0.0178	&	0.0191	\\
9088	&	H$_2$O	&	 Ly$\zeta$	&	0.0112	&	0.0118	&	0.0128	\\
9133	&	H$_2$O	&	 Ly$\eta$	&	0.0089	&	0.0092	&	0.0097	\\
9163	&	H$_2$O	&	 Ly$\theta$	&	0.0018	&	0.0021	&	0.0053	\\
9185	&	H$_2$O	&	 Ly$\iota$	&	0.0369	&	0.0434	&	0.0441	\\
9201	&	H$_2$O	&	 Ly$\kappa$	&	0.0063	&	0.0066	&	0.0352	\\
9213	&	H$_2$O	&	 Ly$\lambda$	&	0.0022	&	0.0316	&	0.0736	\\
9223	&	H$_2$O	&	 Ly$\mu$	&	0.0318	&	0.0880	&	0.0821	\\
9230	&	H$_2$O	&	 Ly$\nu$	&	0.1191	&	0.0873	&	0.0516	\\
9236	&	H$_2$O	&	 Ly$\xi$	&	0.0776	&	0.0350	&	0.0174	\\
\hline
6958	&	CO	&	 Ly$\alpha$ 	&	1.0000	&	1.0000	&	1.0000	\\
8247	&	CO	&	 Ly$\beta$	&	0.1657	&	0.1715	&	0.1790	\\
8698	&	CO	&	 Ly$\gamma$	&	0.0588	&	0.0619	&	0.0661	\\
8907	&	CO	&	 Ly$\delta$	&	0.0292	&	0.0311	&	0.0337	\\
9020	&	CO	&	 Ly$\epsilon$	&	0.0174	&	0.0186	&	0.0205	\\
9088	&	CO	&	 Ly$\zeta$	&	0.0115	&	0.0124	&	0.0139	\\
9133	&	CO	&	 Ly$\eta$	&	0.0091	&	0.0095	&	0.0109	\\
9163	&	CO	&	 Ly$\theta$	&	0.0017	&	0.0035	&	0.0117	\\
9185	&	CO	&	 Ly$\iota$	&	0.0436	&	0.0431	&	0.0533	\\
9201	&	CO	&	 Ly$\kappa$	&	0.0046	&	0.0188	&	0.0625	\\
9213	&	CO	&	 Ly$\lambda$	&	0.0116	&	0.0696	&	0.0870	\\
9223	&	CO	&	 Ly$\mu$	&	0.0772	&	0.0968	&	0.0682	\\
9230	&	CO	&	 Ly$\nu$	&	0.1108	&	0.0586	&	0.0306	\\
9236	&	CO	&	 Ly$\xi$	&	0.0390	&	0.0158	&	0.0073	\\
\hline
\enddata
\tablecomments{SEC line ratios for Fe$^{26+}$ collisions with H, He, H$_2$, N$_2$, H$_2$O, and CO.  These line ratios were obtained by applying the low energy distribution (Equation 10) to MCLZ $n$-resolved cross sections.  For Fe$^{26+}$ collisions with N$_2$, line ratios obtained by applying the SL1 distribution are also presented.  See text for details.}
\end{deluxetable}

\begin{deluxetable}{c c c c c c}
\tabletypesize{\scriptsize}

\tablecaption{Charge Exchange X-ray emission normalized line ratios for Fe $\uppercase\expandafter{\romannumeral 25}$}
\tablewidth{0pt}
\tablehead{
\multicolumn{1}{c}{Photon Energy} &
\multicolumn{1}{c}{Neutral} & 
\multicolumn{1}{c}{Line}&
\multicolumn{1}{c}{}&
\multicolumn{1}{c}{Intensity}&
\multicolumn{1}{c}{}\\
\cline{4-6}
\multicolumn{1}{c}{(eV)} & 
\multicolumn{1}{c}{} & 
\multicolumn{1}{c}{} &
\multicolumn{1}{c}{10 eV/u} &
\multicolumn{1}{c}{100 eV/u} &
\multicolumn{1}{c}{1 keV/u}\\
\multicolumn{1}{c}{} & 
\multicolumn{1}{c}{} & 
\multicolumn{1}{c}{} &
\multicolumn{1}{c}{(43.8 km/s)} &
\multicolumn{1}{c}{(138 km/s)} &
\multicolumn{1}{c}{(438 km/s)}
}
\startdata

6637	&	H	&	 K$\alpha$ f 	&	0.50760	&	0.57366	&	0.65399	\\
6676	&	H	&	 K$\alpha$ i 	&	3.17116	&	3.19627	&	3.23542	\\
6700	&	H	&	 K$\alpha$ r 	&	1.00000	&	1.00000	&	1.00000	\\
7881	&	H	&	 K$\beta$	&	0.17851	&	0.19549	&	0.21538	\\
8296	&	H	&	 K$\gamma$	&	0.06863	&	0.07711	&	0.08720	\\
8487	&	H	&	 K$\delta$	&	0.03492	&	0.03965	&	0.04536	\\
8591	&	H	&	 K$\epsilon$	&	0.02053	&	0.02345	&	0.02700	\\
8653	&	H	&	 K$\zeta$	&	0.01358	&	0.01557	&	0.01803	\\
8694	&	H	&	 K$\eta$	&	0.00964	&	0.01110	&	0.01292	\\
8722	&	H	&	 K$\theta$	&	0.00722	&	0.00835	&	0.00993	\\
8742	&	H	&	 K$\iota$	&	0.00565	&	0.00656	&	0.01133	\\
8757	&	H	&	 K$\kappa$	&	0.00458	&	0.00636	&	0.03785	\\
8768	&	H	&	 K$\lambda$	&	0.00472	&	0.03598	&	0.09074	\\
8777	&	H	&	 K$\mu$	&	0.05162	&	0.09783	&	0.08323	\\
8784	&	H	&	 K$\nu$	&	0.09310	&	0.06116	&	0.03303	\\
8790	&	H	&	 K$\xi$	&	0.03029	&	0.01401	&	0.00640	\\
\hline
6637	&	He	&	 K$\alpha$ f 	&	0.14153	&	0.16738	&	0.20431	\\
6676	&	He	&	 K$\alpha$ i 	&	0.64585	&	0.72242	&	0.81858	\\
6700	&	He	&	 K$\alpha$ r 	&	1.00000	&	1.00000	&	1.00000	\\
7881	&	He	&	 K$\beta$	&	0.20136	&	0.22302	&	0.24269	\\
8296	&	He	&	 K$\gamma$	&	0.07725	&	0.08848	&	0.09933	\\
8487	&	He	&	 K$\delta$	&	0.03880	&	0.04517	&	0.05164	\\
8591	&	He	&	 K$\epsilon$	&	0.02258	&	0.02654	&	0.03072	\\
8653	&	He	&	 K$\zeta$	&	0.01486	&	0.01758	&	0.02063	\\
8694	&	He	&	 K$\eta$	&	0.01055	&	0.01254	&	0.01854	\\
8722	&	He	&	 K$\theta$	&	0.00792	&	0.01160	&	0.06140	\\
8742	&	He	&	 K$\iota$	&	0.01271	&	0.08175	&	0.14309	\\
8757	&	He	&	 K$\kappa$	&	0.12915	&	0.13032	&	0.08885	\\
8768	&	He	&	 K$\lambda$	&	0.05882	&	0.02949	&	0.01449	\\
8777	&	He	&	 K$\mu$	&	0.00177	&	0.00078	&	0.00036	\\
\hline
6637	&	H$_2$	&	 K$\alpha$ f 	&	0.12842	&	0.15124	&	0.18180	\\
6676	&	H$_2$	&	 K$\alpha$ i 	&	0.66993	&	0.74032	&	0.82185	\\
6700	&	H$_2$	&	 K$\alpha$ r 	&	1.00000	&	1.00000	&	1.00000	\\
7881	&	H$_2$	&	 K$\beta$	&	0.18355	&	0.20067	&	0.22028	\\
8296	&	H$_2$	&	 K$\gamma$	&	0.06988	&	0.07857	&	0.08873	\\
8487	&	H$_2$	&	 K$\delta$	&	0.03505	&	0.03992	&	0.04572	\\
8591	&	H$_2$	&	 K$\epsilon$	&	0.02035	&	0.02335	&	0.02699	\\
8653	&	H$_2$	&	 K$\zeta$	&	0.01332	&	0.01536	&	0.01789	\\
8694	&	H$_2$	&	 K$\eta$	&	0.00938	&	0.01087	&	0.01274	\\
8722	&	H$_2$	&	 K$\theta$	&	0.00698	&	0.00813	&	0.00973	\\
8742	&	H$_2$	&	 K$\iota$	&	0.00543	&	0.00635	&	0.01305	\\
8757	&	H$_2$	&	 K$\kappa$	&	0.00437	&	0.00828	&	0.05238	\\
8768	&	H$_2$	&	 K$\lambda$	&	0.00921	&	0.06030	&	0.10122	\\
8777	&	H$_2$	&	 K$\mu$	&	0.08392	&	0.09491	&	0.06742	\\
8784	&	H$_2$	&	 K$\nu$	&	0.06273	&	0.03467	&	0.01756	\\
8790	&	H$_2$	&	 K$\xi$	&	0.01020	&	0.00423	&	0.00196	\\
\hline
6637	&	N$_2$	&	 K$\alpha$ f 	&	0.13302	&	0.15832	&	0.19328	\\
6676	&	N$_2$	&	 K$\alpha$ i 	&	0.68604	&	0.76220	&	0.85266	\\
6700	&	N$_2$	&	 K$\alpha$ r 	&	1.00000	&	1.00000	&	1.00000	\\
7881	&	N$_2$	&	 K$\beta$	&	0.18756	&	0.20578	&	0.22713	\\
8296	&	N$_2$	&	 K$\gamma$	&	0.07190	&	0.08120	&	0.09236	\\
8487	&	N$_2$	&	 K$\delta$	&	0.03617	&	0.04140	&	0.04783	\\
8591	&	N$_2$	&	 K$\epsilon$	&	0.02104	&	0.02427	&	0.02832	\\
8653	&	N$_2$	&	 K$\zeta$	&	0.01378	&	0.01599	&	0.01882	\\
8694	&	N$_2$	&	 K$\eta$	&	0.00971	&	0.01133	&	0.01343	\\
8722	&	N$_2$	&	 K$\theta$	&	0.00723	&	0.00848	&	0.01167	\\
8742	&	N$_2$	&	 K$\iota$	&	0.00563	&	0.00673	&	0.02290	\\
8757	&	N$_2$	&	 K$\kappa$	&	0.00455	&	0.01457	&	0.06817	\\
8768	&	N$_2$	&	 K$\lambda$	&	0.01584	&	0.07397	&	0.10060	\\
8777	&	N$_2$	&	 K$\mu$	&	0.09111	&	0.09026	&	0.06079	\\
8784	&	N$_2$	&	 K$\nu$	&	0.05685	&	0.03048	&	0.01531	\\
8790	&	N$_2$	&	 K$\xi$	&	0.00797	&	0.00358	&	0.00165	\\
\hline
6637	&	H$_2$O	&	 K$\alpha$ f 	&	0.12116	&	0.14483	&	0.17320	\\
6676	&	H$_2$O	&	 K$\alpha$ i 	&	0.68168	&	0.74758	&	0.82379	\\
6700	&	H$_2$O	&	 K$\alpha$ r 	&	1.00000	&	1.00000	&	1.00000	\\
7881	&	H$_2$O	&	 K$\beta$	&	0.17315	&	0.19256	&	0.21120	\\
8296	&	H$_2$O	&	 K$\gamma$	&	0.06488	&	0.07489	&	0.08456	\\
8487	&	H$_2$O	&	 K$\delta$	&	0.03232	&	0.03795	&	0.04346	\\
8591	&	H$_2$O	&	 K$\epsilon$	&	0.01869	&	0.02216	&	0.02559	\\
8653	&	H$_2$O	&	 K$\zeta$	&	0.01219	&	0.01455	&	0.01691	\\
8694	&	H$_2$O	&	 K$\eta$	&	0.00856	&	0.01026	&	0.01200	\\
8722	&	H$_2$O	&	 K$\theta$	&	0.00634	&	0.00764	&	0.00899	\\
8742	&	H$_2$O	&	 K$\iota$	&	0.00491	&	0.00594	&	0.00731	\\
8757	&	H$_2$O	&	 K$\kappa$	&	0.00394	&	0.00480	&	0.01324	\\
8768	&	H$_2$O	&	 K$\lambda$	&	0.00326	&	0.00841	&	0.05183	\\
8777	&	H$_2$O	&	 K$\mu$	&	0.00898	&	0.05743	&	0.08833	\\
8784	&	H$_2$O	&	 K$\nu$	&	0.07650	&	0.08433	&	0.05839	\\
8790	&	H$_2$O	&	 K$\xi$	&	0.06361	&	0.03460	&	0.01728	\\
\hline
6637	&	CO	&	 K$\alpha$ f 	&	0.12945	&	0.15358	&	0.18541	\\
6676	&	CO	&	 K$\alpha$ i 	&	0.69163	&	0.76121	&	0.84516	\\
6700	&	CO	&	 K$\alpha$ r 	&	1.00000	&	1.00000	&	1.00000	\\
7881	&	CO	&	 K$\beta$	&	0.18314	&	0.20065	&	0.22041	\\
8296	&	CO	&	 K$\gamma$	&	0.06987	&	0.07884	&	0.08913	\\
8487	&	CO	&	 K$\delta$	&	0.03508	&	0.04013	&	0.04602	\\
8591	&	CO	&	 K$\epsilon$	&	0.02037	&	0.02349	&	0.02719	\\
8653	&	CO	&	 K$\zeta$	&	0.01332	&	0.01546	&	0.01802	\\
8694	&	CO	&	 K$\eta$	&	0.00937	&	0.01093	&	0.01283	\\
8722	&	CO	&	 K$\theta$	&	0.00696	&	0.00816	&	0.00989	\\
8742	&	CO	&	 K$\iota$	&	0.00540	&	0.00637	&	0.01191	\\
8757	&	CO	&	 K$\kappa$	&	0.00435	&	0.00628	&	0.03770	\\
8768	&	CO	&	 K$\lambda$	&	0.00435	&	0.03437	&	0.08559	\\
8777	&	CO	&	 K$\mu$	&	0.04686	&	0.09086	&	0.08172	\\
8784	&	CO	&	 K$\nu$	&	0.08678	&	0.05925	&	0.03333	\\
8790	&	CO	&	 K$\xi$	&	0.02748	&	0.01305	&	0.00617	\\

\hline
\enddata
\tablecomments{SEC line ratios for Fe$^{25+}$ collisions with H, He, H$_2$, N$_2$, H$_2$O, and CO.   MCLZ $n \ell S$-resolved cross sections were used to obtain these ratios.  See text for details.}
\end{deluxetable}


\begin{thebibliography}{}

\bibitem[Ali et al.(2005)]{ali05} Ali, R., Neill, P. A., Beiersdorfer, P., Harris, C. L., Rakovi\'c, M. J., Wang, J. G.,
   Schultz, D. R., \& Stancil, P. C. 2005, \apj, 629, L125

\bibitem[Bautista \& Kallman(2000)]{bau00} Bautista, M. A. \& Kallman, T. R. 2000, \apj, 544, 581

\bibitem[Bhardwaj et al.(2007)]{bha07} Bhardwaj, A., Elsner, R. F., Gladstone,
   G. R., et al. 2007, Planet. Space Sci., 55, 1135

\bibitem[Butler \& Dalgarno(1980)]{2}
Butler, S.E., \& Dalgarno, A. 1980, ApJ, 241, 838-843

\bibitem[Connerade(1998)]{8}
Connerade, J. P. 1998, \textit{Highly Excited Ions} (Cambridge, Cambridge Univ. Press) 

\bibitem[Cravens(2000)]{cra00} Cravens, T. E. 2000, \apj, 532, L153

\bibitem[Cumbee et al.(2014)]{10}
Cumbee, R. S., Henley, D. B., Stancil, P. C., Shelton, R. L., Nolte, J. L., Wu, Y., \& Schultz, D. R. 2014,  ApJL, 787, L31

\bibitem[Ferland et al.(2013)]{fer13} Ferland, G. J., Porter, R. L., van Hoof, et al. 2013, Rev. Mex. de Astron. y Astrof., 49, 137

\bibitem[Flower(1990)]{flower}
Flower, D. R. 1990, \textit{Molecular Collisions in the Interstellar Medium} (Cambridge, Cambridge Univ. Press)

\bibitem[Foster et al.(2012)]{grat}
Foster, A. R., Smith, R. K., \& Brickhouse, N. S. 2012, ApJ, 756, 128


\bibitem[Herzberg(1950)]{herz}
Herzberg, G. 1950, Spectra of Diatomic Molecules 2nd ed., (New York: Van Nostrand Reinhold)


\bibitem[Janev et al.(1983)]{5}
Janev, R. K., Beli\'{c}, D.S., \& Bransden, B.H. 1983, Phys. Rev. A, 28, 3

\bibitem[Katsonis et al.(1991)]{kats}
Katsonis, K., Maynard, G., \& Janev, R. K. 1991, Physica Scripta, Vol. T37, 80-88

\bibitem[Katsuda et al.(2011)]{suda}
Katsuda, S., Tsunemi, H., Mori, K., et al. 2011, ApJ, 730, 24

\bibitem[Kramida et al.(2014)]{9}
Kramida, A., Ralchenko, Yu., Reader, J., and NIST ASD Team (2014). NIST Atomic Spectra Database (ver. 5.2), \lbrack Online\rbrack. Available: http://physics.nist.gov/asdt \lbrack 2015, June 10\rbrack. National Institute of Standards and Technology, Gaithersburg, MD. 

\bibitem[Krasnopolsky et al.(2004)]{6}
Krasnopolsky, V.  A., Greenwood, J. B., \& Stancil, P. C. 2004, Space Sci. Rev., 113, 271

\bibitem[Lisse et al.(1996)]{lis96} Lisse, C. M., Dennerl, K., Englhauser, J.,
   et al. 1996, Science, 274, 205

\bibitem[Muno et al.(2004)]{mun04} Muno, M. P., et al. 2004, \apj, 613, 326

\bibitem[Olson et al.(1971)]{other}
Olson, R. E., Smith, F. T., \& Bauer, E. 1971, Appl. Opt., 10, 1848

\bibitem[Olson \& Salop(1976)]{3}
Olson, R. E., \& Salop, A. 1976, Phys. Rev. A, 13, 1312

\bibitem[Porquet et al.(2010)]{porquet}
Porquet, D., Dubau, J., \& Grosso, N. 2010, Space Sci. Rev., 157, 103-134

\bibitem[Porter \& Ferland(2007)]{porter}
Porter, R. L., \& Ferland, G. J. 2007, ApJ, 664, 586-595

\bibitem[Rigazio et al.(2002)]{rig02} Rigazio,
  M., Kharchenko, V., \& Dalgarno, A. 2002,  Phys. Rev. A, 66, 064701
  
\bibitem[Schultz et al.(1991)]{schultz}
Schultz, D. R., Meng, L., Reinhold, C. O., \& Olson, R. E. 1991, Physica Scripta, Vol. T37, 89-93

\bibitem[Smith et al.(2014)]{7}
Smith, R. K., Foster,  A. R., Edgar, R.  J., \& Brickhouse, N. S. 2014.  ApJ, 787, 77

\bibitem[Tanaka et al.(1999)]{tan99} Tanaka, Y., Miyaji, T., \& Hassinger, G. 1999, AN, 320, 181

\bibitem[Tauljberg(1986)]{4}
Taulbjerg, K. 1986,  J. Phys. B: At. Mol. Phys., 19, L367

\bibitem[Wargelin et al.(2007)]{war2}
Wargelin, B. J., Beiersdorfer, P., \& Brown, G. V. 2007, Canadian J. Physics, 86, 151

\bibitem[Wargelin et al.(2005)]{1}
Wargelin, B. J., Beiersdorfer, P., Neill, P. A., Olson, R. E., \& Scofield, J. H. 2005, ApJ, 643, 687


\bibitem[Wigner \& Witmer(1928)]{wigner}
Wigner, E., \& Witmer, E. 1928, Z. Phys., 51, 859


\bibitem[Wu et al.(2011)]{wu-stancil}
Wu, Y., Stancil, P. C., Liebermann, H. P., Funke, P., Rai, S. N., Buenker, R. J., Schultz, D. R., Hui, Y., Draganic, I. N., \& Havener, C. C. 2011, Phys. Rev. A, 84, 022711

\end{thebibliography}
\end{document}